\documentclass[a4paper,11pt]{article}

\usepackage{jheppub}

\usepackage[T1]{fontenc}
\usepackage[utf8]{inputenc}
\usepackage{lmodern}
\usepackage{microtype}
\DeclareUnicodeCharacter{2010}{-}

\usepackage{mathtools}
\usepackage{amssymb}
\usepackage{dsfont}
\usepackage{mathrsfs}
\usepackage{tensor}

\usepackage{graphicx}
\usepackage{array}
\usepackage{diagbox}
\usepackage{orcidlink}

\usepackage{xcolor}
\definecolor{dark-blue2}{RGB}{0,114,178}
\definecolor{antiquefuchsia}{rgb}{0.57,0.36,0.51}

\usepackage{hyperref}
\hypersetup{
    pdfencoding=unicode,
    colorlinks=true,
    linkcolor=dark-blue2,
    citecolor=dark-blue2,
    urlcolor=dark-blue2,
    linktoc=section,
    pdftitle={What makes spacetime spin in string theory?},
    pdfauthor={Matilda Delgado, Lorenz Eberhardt, Marija Tomasevic},
    pdfdisplaydoctitle=true,
    pdfstartview=FitH,
    linktocpage=true
}
\usepackage{tocloft}

\newcommand{\ZZ}{\mathbb{Z}}
\newcommand{\RR}{\mathbb{R}}
\newcommand{\TT}{\mathbb{T}}
\newcommand{\RP}{\mathbb{RP}}

\DeclareMathOperator{\Hom}{Hom}
\DeclareMathOperator{\U}{U}
\DeclareMathOperator{\Sq}{Sq}
\renewcommand{\bar}{\overline}
\AtBeginDocument{\renewcommand*{\d}{\mathop{\kern0pt\mathrm{d}}\!{}}}
\newcommand{\svec}[2]{\begin{bsmallmatrix}#1\vphantom{\tilde{\beta}1/2}\\#2\vphantom{\tilde{\beta}1/2}\end{bsmallmatrix}}
\newcommand{\svecsub}[2]{\scalebox{0.6}{$\begin{bsmallmatrix}#1\vphantom{\tilde{\beta}1/2}\\#2\vphantom{\tilde{\beta}1/2}\end{bsmallmatrix}$}}
\newcommand{\FL}{\ensuremath{{\mathrm{F}_{\mathrm{L}}}}}
\newcommand{\FR}{\ensuremath{{\mathrm{F}_{\mathrm{R}}}}}
\newcommand{\Ftot}{\ensuremath{{\mathrm{F}}}}
\newcommand{\Fs}{\ensuremath{{\mathrm{F}_{\mathrm{s}}}}}
\newcommand{\spin}{\ensuremath{\mathrm{spin}}}
\newcommand{\Spin}{\ensuremath{\mathrm{Spin}}}
\newcommand{\pinm}{\ensuremath{\mathrm{pin}^-}}
\newcommand{\pinp}{\ensuremath{\mathrm{pin}^+}}
\newcommand{\Pinm}{\ensuremath{\mathrm{Pin}^-}}
\newcommand{\Pinp}{\ensuremath{\mathrm{Pin}^+}}

\title{What makes spacetime spin in string theory?}

\author{Matilda Delgado$^{a,b}$\orcidlink{0000-0002-8346-4438},}
\author{Lorenz Eberhardt$^c$\orcidlink{0000-0003-1912-2211}, and}
\author{Marija Toma\accent 7 sevi\'c$^d$\orcidlink{0000-0003-4895-2644} }

\affiliation[a]{Jefferson Physical Laboratory, Harvard University\\ 17 Oxford St, Cambridge, MA 02138, United States of America} 
\affiliation[b]{Max-Planck-Institut f\"ur Physik (Werner-Heisenberg-Institut)\\
Boltzmannstr. 8, 85748 Garching, Germany}

\affiliation[c]{Institute for Theoretical Physics,
University of Amsterdam, Amsterdam, 1098XH, NL}

\affiliation[d]{Theoretical Physics Department, CERN, 1211 Geneva 23, Switzerland}
\emailAdd{matildadelgado@fas.harvard.edu}
\emailAdd{l.eberhardt@uva.nl}
\emailAdd{marija.tomasevic@cern.ch}

\abstract{
Type II string theory in the absence of orientifolds requires the target spacetime $X$ to admit a spin structure. We show that this familiar requirement arises directly from the consistency of the worldsheet GSO projection. The obstruction is a mixed global anomaly between $(-1)^{\FL}$ and the target space background data, detected by the spin bordism group $\Omega_3^{\spin}(B\ZZ_2\times X)$. We compute the relevant mixed bordism group and identify the bordism class of the GSO-projected worldsheet theory.
For smooth target spaces, vanishing of the anomaly reduces to the condition that $X$ is spin, while for general orbifolds $[\hat{X}/G]$, $\hat{X}$ has to carry a $G$-equivariant spin structure. We also classify all possible theta angles in the worldsheet theory and show that they correspond to all possible continuous and discrete background fields of the target space theory visible in string perturbation theory.

}

\begin{document}\emergencystretch 3em
\hypersetup{pageanchor=false}
\makeatletter
\let\old@fpheader\@fpheader
\preprint{MPP-2026-109, \; CERN-TH-2026-134}

\maketitle

\flushbottom

\section{Introduction}
One of the early surprises of string theory was that the seemingly simple problem of a string propagating through spacetime gives rise to remarkably powerful consistency conditions. The surprise is especially sharp compared to the worldline QFT description. In it, one merely reorganizes a given QFT in terms of particle trajectories propagating on a fixed spacetime background. The string worldsheet, however, is far less forgiving: the internal consistency conditions of the underlying 2d CFT translate directly into constraints on the admissible spacetime backgrounds. This suggests something far more non-trivial: any 2d CFT satisfying suitable consistency conditions should be viewed as defining an admissible string background, even when no conventional geometric interpretation is available \cite{Gepner:1987vz, Banks:1987cy, Banks:1988yz}.

For the bosonic string, one needs to assume a consistent worldsheet CFT of central charge $c=\bar{c}=26$.\footnote{For spacetime unitarity, one also needs the physical string Hilbert space to carry a positive-definite inner product. For flat target space, this is ensured by the no-ghost theorem \cite{Goddard:1972iy}; for a general matter CFT, it imposes an analogous condition that negative-norm worldsheet states do not survive in the physical spectrum. 
We will set the issue of spacetime unitarity aside in the following.}
For the type 0 string and type II superstring, the worldsheet CFT is an $\mathcal{N}=(1,1)$ SCFT of central charge $c=\bar{c}=15$, while for the heterotic string, it is an $\mathcal{N}=(1,0)$ SCFT of central charge $c=15$, $\bar{c}=26$. The worldsheet SCFT needs to satisfy all the standard CFT bootstrap conditions such as crossing symmetry and modular invariance.

These conditions are enough for the type 0 string to define a consistent string background. The definition of type II strings additionally requires the implementation of the GSO projection that corresponds to a sum over independent spin structures for left- and right-movers \cite{Seiberg:1986by, Gliozzi:1976qd,Alvarez-Gaume:1986ghj}. For a string propagating in a flat target space, the GSO projection eliminates the tachyon and gives rise to spacetime supersymmetry. 
Contrary to the type 0 case, summing over spin structures for both left- and right-movers is not always possible for an $\mathcal{N}=(1,1)$ SCFT, and the obstruction gives rise to additional consistency conditions.
For the heterotic string, the analogous statement is already built into the requirement of a non-anomalous $\mathcal{N}=(1,0)$ worldsheet SCFT: the left-moving fermions valued in $\phi^*TX$ carry a sigma-model anomaly that forces the target space to admit a string structure, i.e.\ in particular a spin structure with $\d H=\frac{1}{2} p_1$ \cite{Witten:1985mj, Witten:1985xe, Killingback:1986rd, Tachikawa:2021mby,Yonekura:2022reu}. The target space condition is thus not an extra input, but part of having the SCFT.

Type 0 theories feature a sum over the joint spin structure of left- and right-movers, which can be viewed as gauging the fermion-parity operator $(-1)^{\Ftot} = (-1)^{\FL + \FR}$.
On the other hand, type II theories are much harder to define, as they require us to sum over left and right sectors \textit{independently}, i.e.\ gauge both $(-1)^\FL$ and $(-1)^\FR$. Beyond the diagonal $(-1)^\Ftot$ already gauged in type 0, this amounts to gauging an additional chiral $\ZZ_2$, say $(-1)^\FL$, which is a priori not part of the datum of an $\mathcal{N}=(1,1)$ SCFT. In fact, the GSO projection can be viewed as gauging the chiral $\ZZ_2$ symmetry \cite{Kaidi:2019tyf}. The distinction between type IIA and type IIB then arises from different discrete theta angles associated with this gauging.

Since the GSO projection is implemented by gauging a chiral $\ZZ_2$ symmetry on the worldsheet, it becomes clear that not every spacetime background admits a consistent GSO projection.\footnote{Throughout this work, we restrict our attention to oriented worldsheets, and therefore do not consider orientifolds. Since orientifold projections involve worldsheet orientation reversal and mix left- and right-moving sectors, the corresponding anomaly problem is qualitatively different and will not be addressed here; we provide further comments in section~\ref{sec:discussion}.} Chiral symmetries are susceptible to global anomalies which obstruct the gauging procedure. Consequently, the requirement of a consistent GSO projection imposes a non-trivial constraint on the target space: admissible spacetime backgrounds must be such that the resulting worldsheet theory is free of anomalies. A question then arises:
is there a natural interpretation of this requirement from the target space point of view?

Assuming that the target space is geometric, we answer this question in the affirmative. For a smooth target space supported by a pure NS-NS background, we find that the consistency of the GSO projection is controlled by a remarkably simple property of the target space:
\begin{quote}
\centering
\emph{Consistency of the chiral GSO projection requires \\ the target space to admit a spin structure.} 
\end{quote}

The reason such a statement can exist at all is that the worldsheet theory is not merely an abstract two-dimensional CFT: its fields include maps $\phi:\Sigma\to X$ into the target space. Consequently, global anomalies may depend not only on the geometric and symmetry data of the worldsheet, but also on topological information pulled back from $X$ through $\phi$. This raises the possibility that anomaly cancellation imposes topological constraints directly on the target space.

At first sight, this result may seem unsurprising since the spacetime theory is known to contain fermions, and on an orientable target with no further gauge fields, these require a spin structure.
However, this conclusion is conventionally reached by first quantizing the string in flat space, reading off the low-energy effective supergravity, and then analyzing which backgrounds it can be consistently placed on. The spin condition then appears as a property of the target space fields. There is a big conceptual leap in this argument when placing the theory obtained from expanding around flat space on non-trivial backgrounds. In particular, this neglects all possible consistency conditions that the full string theory might impose, but which are invisible in the low-energy description.
We should also mention that there are various topological terms in supergravity that make it delicate to place the theory on non-trivial backgrounds \cite{Witten:1996md, Freed:2019sco, Debray:2021vob}.

To make our result precise, we employ bordism theory, which has emerged as the natural framework for classifying global anomalies \cite{Kapustin:2014dxa, Freed:2016rqq, Yonekura:2016wuc, Garcia-Etxebarria:2018ajm}. Roughly speaking, bordism classifies manifolds equipped with the relevant structures according to whether they differ by the boundary of a higher-dimensional manifold. The key observation is that global anomalies are encoded by invertible topological theories in one higher dimension, known as anomaly theories. These assign a phase to closed manifolds equipped with the relevant structure, and this phase depends only on the corresponding bordism class. Anomalies are, therefore, characterized by bordism invariants, which detect whether the partition function can be consistently defined on all backgrounds. Since the anomaly depends only on the topology of the underlying data, it is natural to expect that the consistency of the GSO projection can likewise be characterized by a purely topological property of the target space.

In practice, determining the anomaly proceeds in two steps: first, one
computes the relevant bordism group, whose Pontryagin dual classifies the
possible anomaly theories; second, one identifies which anomaly theory is
realized by the type II worldsheet CFT mapping into a target space $X$. In our case, the relevant bordism group is
\begin{equation}
\Omega_3^{\spin}(B\ZZ_2\times X)\,.
\end{equation}
The degree $3$ reflects the fact that anomalies of a two-dimensional theory are encoded by an invertible theory in one higher dimension, while the superscript `$\spin$' indicates that the relevant 3-manifolds carry a spin structure. The factor $B\ZZ_2$ is the classifying space of the chiral $\ZZ_2$ symmetry associated with the GSO projection, while $X$ denotes the target space. Equivalently, an element of $\Omega_3^{\spin}(B\ZZ_2\times X)$ consists of a spin 3-manifold together with a $\ZZ_2$ gauge field and a map into $X$. More generally, the same framework can be adapted to the orbifold target spaces.

The anomaly itself is characterized by an element of the Pontryagin dual of the bordism group,
\begin{equation}
\mho_3^\spin(B\ZZ_2\times X)=\Hom(\Omega_3^\spin(B\ZZ_2\times X),\U(1))\,.
\end{equation}
It suffices to compute the mixed part of this group, which detects anomalies involving both the chiral GSO symmetry and the topology of the target space. Our first result gives the simple identification
\begin{equation}
\mho_3^\spin(B\ZZ_2 \times X)_\mathrm{mixed}
\cong
e\big(H^2(X,\ZZ_2),H^1(X,\ZZ_2)\big)\,. \label{eq:intro bordism group}
\end{equation}
Here $e(A, B)$ denotes an extension of $B$ by $A$, meaning the anomaly is built from data in both $H^1(X, \ZZ_2)$ and $H^2(X, \ZZ_2)$. The precise extension is specified in \eqref{eq:extension group law}.

To identify which anomaly class is realized by the type II worldsheet theory, we relate these bordism invariants to reduced eta invariants of three-dimensional Dirac operators \cite{Atiyah_Patodi_Singer_1975, Dai:1994kq, Witten:2015aba, Witten:2019bou}. These eta invariants capture the global anomaly of the two-dimensional Majorana--Weyl fermions appearing on the worldsheet and therefore determine the corresponding anomaly theory. We find that the resulting anomaly class is represented by
\begin{equation}
(w_1(X),w_2(X))
\in
\mho_3^\spin(B\ZZ_2 \times X)_\mathrm{mixed}\,,
\end{equation}
where $w_1(X)$ and $w_2(X)$ are the first two Stiefel--Whitney classes of the target space. Consequently, the mixed anomaly vanishes if and only if
\begin{equation}
w_1(X)=w_2(X)=0\,.
\end{equation}
Equivalently, the target space admits a spin structure. Thus, the familiar requirement that type II string backgrounds be spin manifolds emerges directly from the absence of worldsheet GSO anomalies, providing a purely worldsheet derivation of a target space consistency condition. 
This extends the observation in \cite{Freed:1999vc}, where it was shown that, for smooth target spaces, the existence of a spin structure implies the absence of global gauge anomalies. Our discussion extends this to account for gravitational Dai-Freed anomalies on the worldsheet, using the modern language of bordism groups. We explicitly show that this statement is in fact a one-to-one correspondence, and explain how worldsheet anomalies (and associated bordism invariants) are encoded in the Stiefel--Whitney classes of spacetime.

More generally, strings can also propagate on orbifolds. For an orbifold target space $\hat X/G$, the anomaly analysis must take into account not only the topology of $\hat X$, but also the action of the orbifold group $G$. In this case, the relevant topological data are encoded in \textit{equivariant} cohomology, which simply means that the corresponding topological structure is compatible with the action of $G$. For example, an equivariant spin structure is a spin structure on $\hat X$ for which the action of $G$ lifts consistently to the spin bundle.
We find that the GSO anomaly cancellation requires the vanishing of the first two equivariant Stiefel--Whitney classes,
\begin{equation}
w_1^G(\hat X)=w_2^G(\hat X)=0\,.
\end{equation}
Thus, in the general orbifold setting, the familiar requirement that the target space be orientable and spin is replaced by its equivariant counterpart for $\hat{X}$.

Vanishing of the anomaly guarantees that a GSO projection exists, but not
that it is unique: distinct gaugings differ by theta angles, classified by
$\mho_2^\spin(B\ZZ_2\times X)$, respectively, its orbifold analogue, which we also compute. We find that every
such theta angle corresponds to a known piece of target space background
data. The purely gravitational invariants reproduce the distinction between
the type IIA and type IIB projections \cite{Kaidi:2019pzj}, the continuous
theta angles realize the $B$-field background, and the remaining discrete
choices correspond to the spacetime spin structure together with a discrete $\ZZ_2$ gauge bundle.\footnote{For related results on heterotic worldsheets, see~\cite{Yonekura:2022reu}.} In particular, within the class of
backgrounds considered here, the worldsheet analysis accounts for all GSO
projections and reveals no exotic ones.
\bigskip

The paper is organized as follows. In section~\ref{sec:mixed bordism groups}, we derive \eqref{eq:intro bordism group}. Our main technical tool for this is the Smith isomorphism, which we will discuss in detail. In section~\ref{sec:eta invariants}, we discuss eta invariants on the worldsheet and compute the bordism class realized by fermions on the worldsheet. Finally, we apply these technical preparations in section~\ref{sec:target space constraints} to analyze consistency conditions on target spaces. We exemplify our findings with explicit examples and also discuss possible discrete theta angles appearing in the gauging, corresponding to the non-trivial discrete data of the target space background, such as the target space spin structure. We conclude in section~\ref{sec:discussion} with a discussion and an outlook.

\section{Bordism groups} \label{sec:mixed bordism groups}
In this section, we discuss the bordism groups relevant for the GSO projection of type II string theories. For mathematical background on bordism groups, including spin bordism, see e.g.
\cite{Stong1968,RAY_1998,Rudyak1998}. For physicist-friendly introductions
and applications, see e.g. \cite{Garcia-Etxebarria:2018ajm,Debray:2023yrs}. We focus on worldsheet theories in which both a $\ZZ_2$ symmetry and an additional map $f$ into a connected space $X$ are present. The $\ZZ_2$ factor corresponds to $(-1)^{\FL}$, the left-moving worldsheet fermion parity operator. The map $f$ can describe a gauging, in which case $X=BG$ is the classifying space of a finite group $G$, or it can correspond to the target space in string theory, or a combination thereof. We will discuss the precise physical setup in section~\ref{sec:target space constraints}. Such a simultaneous gauging may be obstructed by anomalies, which are classified by the spin bordism group $\Omega_3^{\spin}(B\ZZ_2 \times X)$.

\paragraph{Summary of results.} We consider the mixed part $\Omega_3^\spin(B\ZZ_2 \times X)_\mathrm{mixed}$.
It is defined to be the part of $\Omega_3^\spin(B\ZZ_2 \times X)$ that vanishes whenever we forget either the $\ZZ_2$ gauge bundle or the map into $X$, meaning that
\begin{equation}
    \Omega_3^\spin(B\ZZ_2 \times X)_\mathrm{mixed} :=  \mathrm{ker} \,p_{1 *} \cap \mathrm{ker} \,p_{2 *}\,, \label{eq:definition Omega mixed}
\end{equation}
where $p_1:B\ZZ_2 \times X \to B\ZZ_2$ and $p_2:B\ZZ_2 \times X \to X$ are the projections on the first and second factors; $p_{j *}$ are the corresponding maps induced on the bordism groups. As we will see in the next section, these mixed elements will correspond to mixed anomalies that obstruct a consistent GSO projection.
$\Omega_3^\spin(B\ZZ_2 \times X)_\mathrm{mixed}$ depends on $X$ only through its low-degree mod-2 (co)homology. We will establish this in sections~\ref{sec:smith}--\ref{sec:extension}: the Smith isomorphism reduces it to the reduced bordism group $\widetilde{\Omega}_2^{\pinm}(X)$, whose Atiyah--Hirzebruch spectral sequence (AHSS) receives contributions in the relevant total degree only from $H_1(X,\ZZ_2)$ and $H_2(X,\ZZ_2)$, with the potentially obstructing differential out of $H_3(X,\ZZ_2)$ shown to vanish in section~\ref{sec:ahss}. In fact, since we are only concerned with elements that see both $X$ and the $\ZZ_2$, the only relevant pieces are the mod-2 (co-)homology groups of $X$ in degrees $d\leq 2$. Indeed, the mod-2 cohomology of $B\ZZ_2 = \RP^\infty$ is generated by a single degree 1 element $\alpha$:
\begin{equation}
    H^*( B\ZZ_2,\ZZ_2)= \ZZ_2[\alpha]\,.
\end{equation}
Therefore, the point is to determine which elements of $H^1(X,\ZZ_2)$ and $H^2(X,\ZZ_2)$ combine with the class $[\alpha]$ in a way that leads to a spin bordism invariant. In this section, we will prove that all elements $[a]\in H^1(X,\ZZ_2)$ and $[b]\in H^2(X,\ZZ_2)$ lead to a non-trivial element in $\Omega_3^\spin(B\ZZ_2 \times X)_\mathrm{mixed}$. Moreover, this element will always be of order 2, unless it is associated to an $[a]\in H^1(X,\ZZ_2)$ such that $[a\cup a] \neq  0 \in H^2(X,\ZZ_2)$, in which case it is of order 4. The result of this section is to show that the mixed part of the bordism group is given by a non-trivial extension:
\begin{equation}
\Omega_3^{\spin}(B\ZZ_2 \times X)_\mathrm{mixed}
\cong e\bigl(H_1(X,\ZZ_2),\, H_2(X,\ZZ_2)\bigr)\,,
\end{equation}
where the extension $e(A,B)$ of $B$ by $A$ means that the group fits into a short exact sequence $0 \to A \to e(A,B) \to B \to 0$.
We will actually be mostly interested in the Pontryagin dual of the bordism group, which describes the invertible topological field theories that characterize anomalies:
\begin{equation}
\mho_3^{\spin}(B\ZZ_2 \times X)_\mathrm{mixed}
\cong e\bigl(H^2(X,\ZZ_2),\, H^1(X,\ZZ_2)\bigr)\,,
\end{equation}
where $\mho_3^{\spin}(-):= \Hom(\Omega_3^{\spin}(-), \U(1))$. This extension can be described explicitly as the set $(a,b) \in H^1(X,\ZZ_2)\times H^2(X,\ZZ_2)$, equipped with the group law
\begin{equation}
(a,b)+(a',b')
=
(a+a',\, b+b'+a\cup a')\,. \label{eq:extension group law}
\end{equation}
We spend the rest of this section proving these statements. Readers primarily interested in the physical applications may skip ahead to the next section. 

The remainder of this section is organized as follows. In proving this, the main tool at our disposal is the Smith isomorphism \cite{AndersonBrownPeterson1969} (see also \cite{debray2024smithfibersequenceinvertible} and references therein), which allows us to relate $\Omega_3^\spin(B\ZZ_2 \times X)_\mathrm{mixed}$ to $\widetilde{\Omega}_2^{\pinm} ( X)$. This will be proven in section~\ref{sec:smith}. In section~\ref{sec:ahss}, we will determine $\widetilde{\Omega}_2^{\pinm} ( X)$ up to an extension problem, which we will resolve by studying bordism invariants in section~\ref {sec:extension}. Finally, for completeness, and because we will make use of it in section~\ref{sec:smoothts}, we compute $\Omega_2^{\spin}( B\ZZ_2\times X)$, which is obtained immediately using the Smith isomorphism and the AHSS. 

\paragraph{Mixed bordism group.} The degree three spin bordism groups and their Pontryagin duals have been characterized in full generality in the literature \cite{brumfiel2018pontrjagindual3dimensionalspin}. However, since our case of interest is $\Omega_3^{\spin}(B\ZZ_2 \times X)$, for some space $X$, we adopt a more direct approach tailored to this specific situation. In particular, we focus on those elements of $\Omega_3^{\spin}(B\ZZ_2 \times X)$ that depend non-trivially on both the $B\ZZ_2$ and the $X$ factors. More precisely, let us consider the inclusion and projection maps: 
\begin{align}\label{eq:ipmaps}
    i_1&: B\ZZ_2 \to B\ZZ_2 \times X \,,   & i_2 &: X \to B\ZZ_2 \times X\,,\\
    p_1&: B\ZZ_2 \times X \to B\ZZ_2 \,,   & p_2 &: B\ZZ_2 \times X \to X\,,
\end{align}
and their induced maps $i_{1*}, \,i_{2*},\,p_{1*}, \,p_{2*}$ on bordism groups. Since $B\ZZ_2 \times X$ is a direct product, $p_1 \,\circ \,i_1 = \mathrm{Id}_{B\ZZ_2} $ and  $p_2 \,\circ \,i_2 = \mathrm{Id}_{X} $. By functoriality, this is also true of the pushforward maps. This implies that there are two decompositions of $\Omega_3^{\spin}(B\ZZ_2 \times X)$:
\begin{align}
    \Omega_3^{\spin}(B\ZZ_2 \times X)&\cong i_ {1 * } \Omega_3^{\spin}(B\ZZ_2) \oplus \mathrm{ker} \,p_{1 *}\,,\\
     \Omega_3^{\spin}(B\ZZ_2 \times X)&\cong i_ {2 * } \Omega_3^{\spin}(X) \oplus \mathrm{ker} \,p_{2 *}\,.\label{eq:noice}
\end{align}
For now, everything we have written is also true for degrees $n\neq 3$. What is special about $n=3$ is that $\Omega_3^{\spin}(\mathrm{pt}) =0$.
Since $p_1\circ i_2: X\to B\ZZ_2$ is the constant map (it factors through a point), the induced map $p_{1*}\circ i_{2*}: \Omega_3^{\spin}(X)\to \Omega_3^{\spin}(B\ZZ_2)$ factors through $\Omega_3^{\spin}(\mathrm{pt})$. Since $\Omega_3^{\spin}(\mathrm{pt}) = 0$, we conclude that $p_{1*}\circ i_{2*} = 0$, and by the same argument $p_{2*}\circ i_{1*} = 0$. Therefore $i_{2*}\Omega_3^{\spin}(X)\subset \ker p_{1*}$ and $i_{1*}\Omega_3^{\spin}(B\ZZ_2)\subset \ker p_{2*}$. Since $p_{1*}\circ i_{1*}=\mathrm{id}$, the map $\pi:=i_{1*}\circ p_{1*}$ is an idempotent with image $i_{1*}\Omega_3^{\spin}(B\ZZ_2)$ and kernel $\ker p_{1*}$, realizing the splitting above. As its image lies in $\ker p_{2*}$, the projector $\pi$ restricts to $\ker p_{2*}$, which therefore splits as
\begin{equation}
    \mathrm{ker} \,p_{2 *} = i_{1*}\Omega_3^{\spin} (B\ZZ_2) \, \oplus ( \mathrm{ker} \,p_{1 *} \cap \mathrm{ker} \,p_{2 *})\,.
\end{equation}
Plugging this into \eqref{eq:noice}, we obtain the following decomposition of $\Omega_3^{\spin}(B\ZZ_2 \times X)$ as:
\begin{equation} \label{eq:decomp}
\Omega_3^\spin(B\ZZ_2 \times X)
\cong
i_{2*}\Omega_3^\spin(X)\oplus i_{1*}\Omega_3^\spin(B\ZZ_2)\oplus
( \mathrm{ker} \,p_{1 *} \cap \mathrm{ker} \,p_{2 *})\,,
\end{equation} 
We recognize the last term in this equation as $\Omega_3^\spin(B\ZZ_2 \times X)_\mathrm{mixed}$, see \eqref{eq:definition Omega mixed}.

\subsection{The Smith isomorphism}\label{sec:smith}
In this section, we introduce the principal tool that reduces the computation of $\Omega_3^{\spin}(B\ZZ_2 \times X)_\mathrm{mixed}$ to that of the degree-2 reduced pin$^{-}$ bordism group $\widetilde{\Omega}_2^{\pinm}(X)$. We begin by reviewing the Smith isomorphism in general terms, and then specialize it to the present setting. Finally, we provide a geometric interpretation of the Smith map, which will be useful in our discussion of bordism invariants in section~\ref{sec:extension}.

\paragraph{The Smith isomorphism.} We consider the Smith isomorphism \cite{AndersonBrownPeterson1969} (see also \cite{debray2024smithfibersequenceinvertible} and references therein): 
\begin{equation}\label{eq:smith}
    \Omega_3^{\spin}(B\ZZ_2\times X, X) \cong \Omega_2^{\pinm}(X)\,,
\end{equation}
where the group on the left-hand side is a relative bordism group. For $X= \mathrm{pt}$, we recover the standard Smith isomorphism
\begin{equation} 
\widetilde{\Omega}_3 ^{\spin}(B\ZZ_2) \cong \Omega_3^{\spin}(B\ZZ_2\times \mathrm{pt}, \mathrm{pt}) \cong \Omega_2^{\pinm}(\mathrm{pt})\,.
\end{equation}
This isomorphism follows from the underlying homotopy equivalence of spectra \begin{equation}\label{eq:spectra}
    MT\Pinm\simeq  M\Pinp \simeq M\mathrm{O}(1) \wedge M\Spin\,, 
\end{equation} 
by smashing with $X$ \cite{AndersonBrownPeterson1969} (see Corollary 1.4.12 in \cite{hertl}). The first equivalence on \eqref{eq:spectra} follows from the fact that a pin$^{-}$ structure on the tangent bundle is equivalent to a pin$^+$ structure on the stable normal bundle \cite{cobordismpin}. Since the Pontryagin-Thom theorem \cite{Thom1954QuelquesPG} identifies bordism groups with homotopy groups of Thom spectra defined using structures on the stable normal bundle, manifolds with a tangential pin$^{-}$ structure are represented by the Thom spectrum $M\Pinp$. Thus the bordism theory encoded by $MT\Pinm$ is equivalently encoded by $M\Pinp$.\footnote{This subtlety does not arise for e.g. spin bordism since a spin structure on the tangent bundle determines one on the stable normal bundle.
} The second equivalence in \eqref{eq:spectra} is the spectral form of the Smith construction \cite{AndersonBrownPeterson1969}. It identifies a stable normal $\pinp$ structure with spin data together with a real line bundle, the latter being encoded by the factor $M\mathrm{O}(1)$, which is the Thom spectrum built from the universal real line bundle over $B\mathbb{Z}_2$.

The relative bordism group on the left-hand side of \eqref{eq:smith} fits in a long exact sequence: 
\begin{equation}
    \cdots \xrightarrow[]{\partial}\Omega_3^{\spin}(X) \xrightarrow[]{i_{2 *}}\Omega_3^{\spin}(B\ZZ_2 \times X) \xrightarrow[]{}\Omega_3^{\spin}(B\ZZ_2\times X, X) \xrightarrow[]{} \cdots\,,
\end{equation}
where $i_{2*}$ is the push-forward of $i_2$ on bordism groups \eqref{eq:ipmaps}. As shown in the previous section, $i_{2*}$ is split injective, so the long exact sequence breaks to: 
\begin{equation}
   0 \to\Omega_3^{\spin}(X) \xrightarrow[]{i_{2 *}}\Omega_3^{\spin}(B\ZZ_2 \times X) \xrightarrow[]{}\Omega_3^{\spin}(B\ZZ_2\times X, X) \xrightarrow[]{}0\,.
\end{equation}
Therefore, 
\begin{equation}
    \Omega_3^{\spin}(B\ZZ_2\times X, X) \cong \frac{\Omega_3^{\spin}(B\ZZ_2 \times X) }{i_{2 *} \Omega_3^{\spin}(X)}\cong \mathrm{ker} \,p_{2*}\,,
\end{equation}
where in the second equation, we have used \eqref{eq:noice}. Using \eqref{eq:decomp}, we have therefore proven that there are isomorphisms:
\begin{equation}\label{eq:smith splitting}
    \Omega_2^{\pinm}(X) \cong \mathrm{ker} \,p_{2*} \cong   i_{1*}\Omega_3^{\spin}(B\ZZ_2) \oplus \Omega_3^{\spin}(B\ZZ_2 \times X)_\mathrm{mixed}\,.
\end{equation} 
For $X=\mathrm{pt}$, the mixed part is absent and therefore $i_{1*}\Omega_3^{\spin}(B\ZZ_2) \cong \Omega_2^{\pinm}(\mathrm{pt})$.
Now, by definition, the reduced bordism group is 
\begin{equation}
    \widetilde{\Omega}_2^{\pinm}(X)= \mathrm{ker}\left( \Omega_2^{\pinm}(X) \to \Omega_2^{\pinm}(\mathrm{pt})\right)\,.
\end{equation}
Therefore,
\begin{equation}    \widetilde{\Omega}_2^{\pinm}(X) \cong \Omega_3^{\spin}(B\ZZ_2 \times X)_\mathrm{mixed} := \mathrm{ker} \,p_{1 *} \cap \mathrm{ker} \,p_{2 *}\,.
\end{equation}
With this in hand, we can now turn to the computation of $\widetilde{\Omega}_2^{\pinm} ( X)$ using the AHSS. Before that, we sketch how classes in $\Omega_3^{\spin}(B\ZZ_2\times X)$ can be mapped to classes in $\Omega_2^{\pinm} (X)$.

\paragraph{Geometric construction of the Smith map.} An element of $\Omega_{3}^{\spin}( B\ZZ_2 \times X)$ is represented by a triple $(M,f_3)$ where $M$ is a closed spin $3$–manifold, $f_3 : M \to B\ZZ_2 \times X$. Recall our space $B\ZZ_2 \times X$, and let $\sigma\to B\ZZ_2 \simeq \RP^\infty$ be the tautological real line bundle. One can pull back this line bundle to a real line bundle over $B\ZZ_2 \times X$ using the map $p_1$ defined in \eqref{eq:ipmaps}. We still call this line bundle $\sigma$. The Smith homomorphism 
\begin{equation}\label{eq:smhomo}
    \mathrm{sm}:\Omega_3^{\spin}(B\ZZ_2 \times X)\to \Omega_2^{\pinm} (X)
\end{equation} induced by the isomorphism \eqref{eq:smith splitting} can be constructed geometrically as follows.  Choose a transverse section $s$ of the real line bundle $L = f_3^*(\sigma)$. Its zero locus $ \Sigma := s^{-1}(0) \subset M $ is a closed smooth surface of dimension $2$. In other words, $\Sigma$ is the surface that is Poincar\'e dual to $f_3^*(\alpha)$, the pullback on $M$ of the generator of $H^1(B\ZZ_2, \ZZ_2)$. 

We now restrict the tangent bundle of $M$ to $\Sigma$. We can split $TM|_\Sigma \cong T\Sigma \oplus L$, since the normal bundle is isomorphic to $L$
because $\Sigma$ is the zero locus of a transverse section of $L$. Therefore, since $M$ is oriented and spin, $w_1(TM)=0$ and $w_2(TM)=0$. Using Whitney sum formulas:
\begin{equation}
0=w_1(TM|_\Sigma)=w_1(T\Sigma)+w_1(L)\quad \Longrightarrow\quad w_1(\det(T \Sigma))=w_1(T\Sigma)=w_1(L)\,.
\end{equation}
Since real line bundles over $\Sigma$ are classified by $H^1(\Sigma,\ZZ_2)$, this equality implies that $L$ is isomorphic to the orientation line bundle $\det(T\Sigma)$. For $w_2$:
\begin{equation}
0=w_2(TM|_\Sigma)=w_2(T\Sigma)+w_1(T\Sigma)\cup w_1(L)+w_2(L)\,.
\end{equation}
Because $L$ is a line bundle, $w_2(L)=0$, hence
\begin{equation}
w_2(T\Sigma)=w_1(T\Sigma)\cup w_1(L)=w_1(T\Sigma)^2\,.
\end{equation}
The existence condition for a $\pinm$ structure on a rank-2 bundle $E$ is
\begin{equation}
w_2(E)+w_1(E)^2=0\,.
\end{equation}
Therefore, we have shown that $T\Sigma$ admits a $\pinm$ structure canonically induced from the spin structure on $M$ together with the identification of the normal bundle with $L$. In particular, $\Sigma$ becomes a closed $\pinm$ surface.  The composition $g := p_2 \circ f_3|_\Sigma : \Sigma \to X$ went along for the ride, and we have thus obtained an element of $\Omega_2^{\pinm } (X)$. From the discussion above, it is clear that elements of $i_{2*}\Omega_3^{\spin}(X)$ are in the kernel of \eqref{eq:smhomo}; this can be seen geometrically from the fact that the line bundle $L$ is trivial for these classes. 

\subsection{AHSS computation of \texorpdfstring{$\widetilde{\Omega}_2^{\pinm}(X)$}{Omega2pin-(X)}}\label{sec:ahss}

We now turn to the computation of the reduced \pinm-bordism group $\widetilde{\Omega}_2^{\pinm}(X)$ using the AHSS.\footnote{For an introduction to the AHSS for physicists, see e.g. \cite{Garcia-Etxebarria:2018ajm}.} The spectral sequence arises from the fibration $\mathrm{pt} \longrightarrow X \longrightarrow X$,
and has $E^2$-page
\begin{equation}
E^2_{p,q} \;=\; H_p\!\left(X,\,\Omega_q^{\pinm}(\mathrm{pt})\right)
\;\Longrightarrow\;
\Omega_{p+q}^{\pinm}(X)\,.
\end{equation}
Our goal is therefore to analyze the relevant low-degree terms and differentials 
in total degree $2$. We only need the low-degree bordism groups of a point:
\begin{equation}
\Omega_0^{\pinm}(\mathrm{pt})\cong \ZZ_2,\qquad
\Omega_1^{\pinm}(\mathrm{pt})\cong \ZZ_2,\qquad
\Omega_2^{\pinm}(\mathrm{pt})\cong \ZZ_8\,.
\end{equation}
For total degree $p+q=2$, the $E^2$ page has only
\begin{equation}
E^2_{0,2}=H_0(X,\ZZ_8)\cong \ZZ_8,\qquad
E^2_{1,1}=H_1(X,\ZZ_2),\qquad
E^2_{2,0}=H_2(X,\ZZ_2)\,.
\end{equation}
The part of the $E^2$ page which is relevant for our purposes is shown here:
\begin{equation} 
    \begin{tabular}{c|cccc}
    2 & $\ZZ_8$ & $H_1(X,\ZZ_8)$ & $H_2(X,\ZZ_8)$ & $H_3(X,\ZZ_8)$ \\
    1 & $\ZZ_2$ & $H_1(X,\ZZ_2)$ & $H_2(X,\ZZ_2)$ & $H_3(X,\ZZ_2)$ \\
    0 & $\ZZ_2$ & $H_1(X,\ZZ_2)$ & $H_2(X,\ZZ_2)$ & $H_3(X,\ZZ_2)$ \\
    \hline
    \diagbox[dir=NE,height=1.4\line]{$q$}{$p$} & 0 & 1 & 2 & 3  \\
    \end{tabular}
\end{equation}
We focus on the reduced bordism groups, for which the zeroth column vanishes. The only potentially non-trivial differential influencing $\Omega_2^{\pinm}(X)$ is
\begin{equation}
    d_2  :E^2_{3,0}=H_3(X,\ZZ_2)\longrightarrow E^2_{1,1}=H_1(X,\ZZ_2)\,.
\end{equation}
The dual differential
\begin{equation}
d_2^\vee:H^1(X,\ZZ_2)\longrightarrow H^3(X,\ZZ_2)
\end{equation}
is a stable mod-$2$ cohomology operation of degree $2$. This is always true for the differentials on the second page of the AHSS. Hence $d_2^\vee$ is either $0$ or $\Sq^2$. Since $\Sq^2$ vanishes on $H^1(-,\ZZ_2)$, it follows that $d_2^\vee=0$ and therefore $d_2=0$. The spectral sequence thus collapses in this range, $E_{p,q}^\infty=E_{p,q}^2$ for $p+q \le 2$.

We obtain a short exact sequence for the reduced bordism group corresponding to the extension problem of the AHSS:
\begin{equation} \label{eq:sesahss}
0\longrightarrow H_1(X,\ZZ_2)\longrightarrow \widetilde{\Omega}_2^{\pinm}(X)\longrightarrow H_2(X,\ZZ_2)\longrightarrow 0\,.
\end{equation}
After dualization, the arrows turn around and we have
\begin{equation} \label{eq:sesahss dual}
0\longrightarrow H^2(X,\ZZ_2)\longrightarrow \widetilde{\mho}_2^{\pinm}(X)\longrightarrow H^1(X,\ZZ_2)\longrightarrow 0\,.
\end{equation}
Thus, the relevant group we are after is an extension of $H^2(X,\ZZ_2)$ and $H^1(X,\ZZ_2)$,
\begin{equation}\label{eq:ext}
\widetilde{\mho}_2^{\pinm}(X)\cong e\big(H^2(X,\ZZ_2),H^1(X,\ZZ_2)\big)\,.
\end{equation}
The AHSS gives the existence of this extension, but not its group law. The remaining job is to determine the extension class.

\subsection{Solving extension problems}\label{sec:extension}
The AHSS in the previous section led to the short exact sequence \eqref{eq:sesahss}.
Resolving the extension problem in \eqref{eq:ext} amounts to determining which pairs $(a,b)\in H^1(X,\ZZ_2)\times H^2(X,\ZZ_2)$ combine non-trivially to form a $\ZZ_4$ summand in $\widetilde{\mho}_2^{\pinm}(X)$, rather than contributing independent $\ZZ_2$ summands. To analyze this, we identify the bordism invariants of $\widetilde{\Omega}_2^{\pinm}(X)$, namely those associated to closed \pinm surfaces equipped with a map to $X$. Let $(\Sigma,f)$ be a closed \pinm surface with $f:\Sigma\to X$. A class $b\in H^2(X,\ZZ_2)$ defines the ordinary cohomological bordism invariant
\begin{equation}\label{eq:deg2inv}
\widetilde{I}_b(\Sigma,f)=\langle f^*b,[\Sigma]\rangle \in \ZZ_2\,,
\end{equation}
where $[\Sigma]$ denotes the fundamental class.
This describes the subgroup $H^2(X,\ZZ_2)\subset \widetilde{\mho}_2^{\pinm}(X)$. To understand the quotient by this subgroup, let $a\in H^1(X,\ZZ_2)$. We pull back $a$ along $f$, giving a class $f^*a\in H^1(\Sigma,\ZZ_2)$. There is a one-to-one correspondence between \pinm structures on $\Sigma$ and quadratic refinements of the mod-$2$ intersection form (section 3 in \cite{Kirby_Taylor_1991}), characterized by
\begin{equation} 
Q_\Sigma:H^1(\Sigma,\ZZ_2)\to \ZZ_4\quad \text{with}\quad Q_\Sigma(a+a')=Q_\Sigma(a)+Q_\Sigma(a')+2\langle a \cup a', [\Sigma] \rangle\,.\label{eq:quadraf}
\end{equation}
The value $Q_\Sigma(f^*a)$ is therefore determined entirely by the $\pinm$ structure on $\Sigma$ together with the class $a\in H^1(X,\ZZ_2)$, or equivalently the induced $\ZZ_2$ gauge field $f^*a\in H^1(\Sigma,\ZZ_2)$. From the results of \cite{Kirby_Taylor_1991, Brown1972Kervaire}, it follows that we can define a $\ZZ_4$-valued $\pinm$ bordism invariant,
\begin{equation}\label{eq:bordinv2}
\widetilde J_a(\Sigma,f):=Q_\Sigma(f^*a)\in \ZZ_4\,.
\end{equation}

One can see that the quadratic refinement property \eqref{eq:quadraf} of this bordism invariant resolves the extension problem above. Indeed, for
$a,a'\in H^1(X,\ZZ_2)$ one has 
\begin{equation} 
\widetilde J_{a+a'}(\Sigma,f)=Q_\Sigma(f^*a+f^*a') =Q_\Sigma(f^*a)+Q_\Sigma(f^*a')
   +2\, \big\langle f^*(a \cup a'),[\Sigma] \big\rangle\,.
\end{equation}
Using the definition of \eqref{eq:deg2inv} gives
\begin{equation}
\widetilde J_a(\Sigma,f)+\widetilde J_{a'}(\Sigma,f)=\widetilde J_{a+a'}(\Sigma,f)+
2\,\widetilde{I}_{a\cup a'}
\quad\text{in }\ZZ_4\,. \label{eq:J twisted group law}
\end{equation}

The invariants $\widetilde I_b$ and $\widetilde J_a$ exhaust the invariants of $\widetilde{\mho}_2^{\pinm}(X)$, and together they must realize its group law. The property \eqref{eq:J twisted group law} tells us that the group law is given by \eqref{eq:extension group law}.
\subsection{The second bordism group} \label{sec:Omega2}
For completeness, we also compute parts of the group $\Omega_2^\spin(B\ZZ_2 \times X)$ that physically will label different possible discrete theta angles. In this case, $\Omega_2^\spin(\mathrm{pt}) \cong \ZZ_2$ is non-trivial, and thus the splitting \eqref{eq:decomp} does not necessarily hold. Instead, the best we can do is to split
\begin{equation} 
\Omega_2^\spin(B\ZZ_2 \times X) \cong i_{2*} \Omega_2^\spin(X) \oplus \ker p_{2*}\,,
\end{equation}
and the Smith isomorphism implies
\begin{equation} 
\ker p_{2*} \cong \Omega_1^{\pinm}(X) \cong \ZZ_2 \oplus H_1(X,\ZZ_2)\,,
\end{equation}
with the last statement following from the AHSS as in section~\ref{sec:ahss}, but without any extension problems.

It remains to describe the first summand $i_{2*}\Omega_2^\spin(X)\cong\Omega_2^\spin(X)$. As with $\widetilde{\mho}_2^{\pinm}(X)$, its reduced Pontryagin dual $\widetilde{\mho}_2^\spin(X)$ is an extension of $H^1(X,\ZZ_2)$ by $H^2(X,\mathrm{U}(1))$, with group law
\begin{equation} 
(a,b) + (a',b')=(a+a',b+b'+\tfrac{1}{2} a \cup a')\,, \label{eq:Gu--Wen group law}
\end{equation}
where we identified $\mathrm{U}(1) \cong \RR/\ZZ$. This is the Gu--Wen extension \cite{Gu:2012ib, Bhardwaj:2016clt}.
We can thus write the full answer as follows,
\begin{equation} 
\mho_2^\spin(B\ZZ_2 \times X) \cong e(H^2(X,\mathrm{U}(1)),H^1(X,\ZZ_2)) \oplus H^1(X,\ZZ_2) \oplus \ZZ_2^2\,. \label{eq:mho 2}
\end{equation}
The bordism invariants are straightforward to write down. Let $Q_\Sigma^{\mathrm{L}/\mathrm{R}}:H^1(\Sigma,\ZZ_2) \to \ZZ_2$ be the quadratic refinements of the intersection pairing associated to the left- and right- spin structure. Both satisfy
\begin{equation} 
Q_\Sigma(a+a')=Q_\Sigma(a)+Q_\Sigma(a')+\langle a \cup a',[\Sigma] \rangle \in \ZZ_2\,,
\end{equation} 
analogously to \eqref{eq:quadraf}.
Let $\mathrm{Arf}^{\mathrm{L}/\mathrm{R}}$ be the Arf invariants corresponding to the left- and right-moving spin structure (that equal 0 for even spin structures and 1 for odd spin structures). We then have the following invariants. For $f:\Sigma \to X$, $b \in H^2(X,\U(1))$ with $\U(1)=\RR/\ZZ$ and $a^\mathrm{L} \in H^1(X,\ZZ_2)$,
\begin{equation} 
\exp\big(2\pi i \langle f^*b,[\Sigma] \rangle \big)\,, \quad \exp\big(\pi i Q_\Sigma^\mathrm{L}(a^\mathrm{L})\big)\,, \quad \exp(\pi i \mathrm{Arf}^\mathrm{L})\,. \label{eq:mho2 bordism invariants}
\end{equation}
The last two bordism invariants also exist with $\mathrm{L} \to \mathrm{R}$.

\section{Eta invariants for fermions in two dimensions} \label{sec:eta invariants}

After laying the groundwork of computing the relevant anomaly groups, we will turn to eta invariants. Fermion anomalies that appear in the superstring are captured by eta invariants, which define elements of the bordism group that we determined in the previous section. The main result will be the identification of a particular (reduced) eta invariant with a bordism class \eqref{eq:eta-to-pin-invariants}. We will use this result in the next section~\ref{sec:target space constraints} to study target space constraints in type II string theory.

\subsection{Eta invariants}\label{sec:eta}

\paragraph{Eta invariants and fermion anomalies.}
The Dai--Freed theorem relates fermion anomalies in $d$ dimensions to
eta invariants in one higher dimension
\cite{Dai:1994kq,Witten:2015aba,Yonekura:2016wuc,Witten:2019bou}.
Consider fermions on a $d$-dimensional manifold $Y$, whose kinetic operator is
a Dirac operator $\mathcal{D}^E_Y$, where the fermions can take values in a spinor bundle twisted by a vector bundle $E$, such as the gauge bundle for a symmetry group $G$. The fermion partition function is generally
not a complex number, but rather a section of an anomaly line bundle over the
space of background fields. Therefore, its phase is not intrinsically defined
on a single background. What is intrinsically defined is the holonomy of the anomaly line bundle around a closed loop.

Let $Y_1$ and $Y_2$ be two $d$-dimensional backgrounds, and let $M$ be a
$(d+1)$-dimensional manifold interpolating between them,
\begin{equation}
    \partial M = Y_1 \sqcup \overline{Y}_2\,.
\end{equation}
The Dirac operator $\mathcal{D}^E_Y$ on the boundary determines a
$(d+1)$-dimensional Dirac operator $\mathcal{D}^E_M$ on $M$, with APS boundary
conditions. The Dai--Freed theorem then states that the relative anomaly phase is given by the exponentiated eta invariant of $\mathcal{D}^E_M$:
\begin{equation}
    \frac{Z_{\mathrm{fermion}}(Y_1)}
         {Z_{\mathrm{fermion}}(Y_2)}
    =
    \exp\left(-2\pi i\,\eta(\mathcal{D}^E_M)\right)\,.
\end{equation}
Equivalently,
at the level of phases,
\begin{equation}
    \arg Z_{\mathrm{fermion}}(Y_1)
    -
    \arg Z_{\mathrm{fermion}}(Y_2)
    =
    -2\pi\,\eta(\mathcal{D}^E_M)
    \mod 2\pi\,.\label{eq:complexfermionanom}
\end{equation}
Here, $\eta$ is defined as the $\zeta$-function regularized sum of the sign of Dirac eigenvalues of $\mathcal{D}_M^E$,
\begin{equation}
\eta(\mathcal{D}_M^E):=\frac{1}{2}\Bigg(\sum_{\lambda \ne 0} \frac{\mathrm{sign}(\lambda)}{|\lambda|^s} +\dim \ker \mathcal{D}_M^E \Bigg)\Bigg|_{s=0}\,,
\end{equation}
where the factor of $1/2$ is the standard APS normalization: when a single eigenvalue crosses zero, the regularized quantity in parentheses changes by $2$, so the reduced eta invariant changes by $1$. Anomaly cancellation means that this phase is trivial for every closed $(d+1)$-dimensional background obtained in this way. In particular, if the interpolation is closed by identifying $Y_1$ with $Y_2$, then $M$ is like a mapping torus with non-trivial topology, and the condition is
\begin{equation}
    \exp\left(-2\pi i\,\eta(\mathcal{D}^E_M)\right)=1
\end{equation}
for every such mapping torus $M$. More generally, the same condition should
hold on all closed $(d+1)$-dimensional backgrounds compatible with the
structure of the theory.

\paragraph{Eta invariants as bordism invariants.}
The previous discussion suggests that $\eta(\mathcal{D}^E_M)$ should be
related to bordism invariants of the $(d+1)$-dimensional background $M$. This
relation is made precise by the APS index theorem
\cite{Atiyah_Patodi_Singer_1975}. Let $W$ be a $(d+2)$-dimensional manifold
with boundary
\begin{equation}
    \partial W = M\,.
\end{equation}
Then the APS index theorem gives
\begin{equation}
    \mathrm{index}(\mathcal{D}^E_W)
    =
    \int_W \widehat{A}(R)\,\mathrm{ch}(E)
    -
    \eta(\mathcal{D}^E_M)\,, \label{eq:APS-index-theorem}
\end{equation}
where $\mathrm{ch}(E)$ is the Chern character of the vector bundle $E$.
Depending on conventions, the sign of the boundary eta term may be reversed;
what is important is that the index is an integer, so the eta invariant is
determined modulo one by the bulk characteristic class. 

The index is by definition integer and thus
\begin{equation}\label{eq:eta}
    \eta(\mathcal{D}^E_M)
    =
    \int_W \widehat{A}(R)\, \mathrm{ch}(E)
    \mod 1\,.
\end{equation}
The right-hand side captures the local gravitational and gauge anomalies of the $d$-dimensional theory. Since $\eta(\mathcal{D}^E_M)$ does not vanish for a null-bordant manifold, it is not by
itself, a bordism invariant unless the local anomaly vanishes. 

We will define below certain reduced eta invariants. These are combinations of eta invariants, built in a way that cancels the contribution coming from local anomalies. For example, when $E$ is a $G$-bundle with $G$ a finite symmetry group, then $\mathrm{ch}(E)-\mathrm{rank}(E)$ vanishes automatically and only the gravitational
piece in the right-hand side of \eqref{eq:eta} remains. In that case, we can consider the reduced eta invariant defined by
\begin{equation}
    \widetilde{\eta}(\mathcal{D}^E_M)
    :=
    \eta(\mathcal{D}^E_M)
    -
    \eta(\mathcal{D}^{0}_M)\,,
\end{equation}
where $\mathcal{D}^E_M$ denotes the Dirac operator twisted by the flat vector bundle $E$, and $\mathcal{D}^{0}_M$ denotes the same Dirac operator with the
trivial bundle of the same rank. Since the subtraction removes the purely gravitational
piece in \eqref{eq:eta}, the reduced invariant depends only on the spin
manifold $M$ together with its classifying map of the vector bundle $E$ into $BG$. In particular, for $G$ a finite group,
\begin{equation}
    \widetilde{\eta}
    :
    \Omega_{d+1}^{\spin}(BG)
    \longrightarrow
    \mathrm{U}(1) =\RR/\ZZ
\end{equation}
defines a bordism invariant, i.e.\ $\widetilde{\eta} \in \mho^\spin_{d+1}(BG)$. 

\paragraph{Eta invariants in three dimensions.}
For the case of interest, $Y$ is the two-dimensional worldsheet, and
$\mathcal{D}^E_M$ is a three-dimensional Dirac operator, possibly twisted by a vector bundle $E$.

Since three dimensions do not admit Weyl fermions, the
three-dimensional Dirac operator naturally packages the data associated with
both two-dimensional chiralities. The reduced eta invariants therefore give
bordism invariants of
\begin{equation}
    \Omega_{3}^{\spin}(X)\,,
\end{equation}
where $X$ is the classifying space of the structure of the vector bundle $E$.
These invariants capture the global part of the fermion anomaly after the
purely gravitational contribution has been removed. 

For real vector bundles $E$, there can be an additional simplification, since it makes sense to consider Majorana fermions. On a spin 3-manifold $M$, a complex chiral fermion has
determinant line anomaly \eqref{eq:complexfermionanom}
\begin{equation}
    \exp\left(-2\pi i\,\eta(\mathcal{D}^E_M)\right)\,.
\end{equation}
Instead, the partition function of a real chiral fermion features the Pfaffian rather than a determinant. Since
\begin{equation}
     \det \mathcal{D} = {\rm Pf}(\mathcal{D})^2\,,
\end{equation}
the anomaly of the Pfaffian is a square root of the anomaly of the determinant. Thus a single Majorana--Weyl fermion contributes \cite{Witten:2015aba, Witten:2019bou}
\begin{equation}
     \exp\left(-\pi i\,\eta(\mathcal{D}^E_M)\right)\,.
\end{equation}
Likewise, since $E$ is real, the Dirac operator $\mathcal{D}_W^E$ on the bounding 4-manifold inherits a quaternionic (pseudoreal) structure. Its kernel is therefore a quaternionic vector space, and the index in \eqref{eq:APS-index-theorem} is even \cite{Tachikawa:2018njr}, meaning that
\begin{equation} 
\frac{1}{2}\eta(\mathcal{D}_M^E)=\frac{1}{2}\int_W \widehat{A}(R)\, \mathrm{ch}(E)
    \mod 1\,. \label{eq:1/2-APS-index-theorem}
\end{equation}
After passing to reduced eta invariants, this allows one to construct more powerful bordism invariants for real vector bundles.

\subsection{Computation of the anomaly} \label{subsec:computation anomaly}
In section~\ref{sec:mixed bordism groups}, we determined the Pontryagin dual of the bordism group
\begin{equation}
     \mho_3^{\spin}(B\ZZ_2 \times X)_\mathrm{mixed}\cong \widetilde{\mho}_2^{\pinm}(X) \cong e(H^2(X,\ZZ_2),H^1(X,\ZZ_2))\,,
\end{equation}
with bordism invariants described in \eqref{eq:deg2inv} and \eqref{eq:bordinv2}.
We now want to connect these bordism invariants to (reduced) eta invariants of 3d Dirac operators, corresponding to the anomaly of two-dimensional Majorana--Weyl
fermions living on the type II worldsheet. 

\paragraph{Reduced eta invariant.} Let $E\to M$ be a real orthogonal vector bundle of rank $r$. In the sigma model application below, $E$ will be the pullback of the target space tangent bundle via the bosonic embedding map, or its equivariant version.

We therefore assume that $E=f^*V$ is the pullback of a real vector bundle $V\to X$ along a map $f:M\to X$. For a sigma model with a smooth target space, $X$ is the target spacetime, $V=TX$, and $f$ is the bosonic embedding map, extended to $M$. The universal case is $X=B\mathrm{O}(r)$, with $V$ the universal bundle and $f$ the classifying map of $E$; any other choice reduces to it by composing $f$ with the classifying map $X\to B\mathrm{O}(r)$ of $V$. We will keep $X$ arbitrary for now.

Suppose we are given the data of a bordism class in $\Omega_3^\spin(B\ZZ_2 \times X)$, i.e.\ a closed spin 3-manifold $M$, a map
$f:M\to X$, and a class
\begin{equation}
     a\in H^1(M,\ZZ_2)\,.
\end{equation}
We denote by $L_a\to M$ the associated real line bundle. We consider the anomaly of a
two-dimensional real fermion valued in $E$, coupled to the
$\ZZ_2$-background $a$ for the $(-1)^{\FL}$ operator. The left-moving fermion thus takes values in $E \otimes L_a$, while the right-moving fermion takes values in $E$. The anomaly is determined by $\eta(\mathcal{D}_M^{E \otimes L_a})-\eta(\mathcal{D}_M^{E})$, accounting for the fact that the anomaly of the right-movers contributes with opposite sign.

In order to determine the mixed anomaly between $(-1)^{\FL}$ and the
$X$-background, we should remove the purely $(-1)^{\FL}$ $\ZZ_8$ anomaly. Equivalently,
we can replace $E$ by the virtual rank-zero bundle $E-\mathrm{rank}(E)\epsilon$, where $\epsilon$ denotes the trivial real line bundle.
Thus we define the reduced eta invariant
\begin{equation} 
        \widetilde\eta_E(M,a)
        :=\eta(\mathcal{D}^{E\otimes L_a}_M)-\eta(\mathcal{D}^{E}_M)  -\mathrm{rank}(E)
        \big(\eta(\mathcal{D}^{L_a}_M)-\eta(\mathcal{D}_M^0)\big)\,. \label{eq:reduced-eta-E}
\end{equation}
The second term on the right-hand side subtracts the part of the eta invariant that is insensitive to $(-1)^{\FL}$ and the last term subtracts the purely $(-1)^{\FL}$ part. $\widetilde\eta_E(M,a)$ vanishes when either $E$ is trivial or when $L_a$ is trivial and thus indeed measures the mixed anomaly. The local terms on the RHS of \eqref{eq:1/2-APS-index-theorem} cancel out between the four terms and thus 
\begin{equation} 
        \frac{1}{2}\widetilde{\eta}_E:\Omega_3^{\spin}(B\ZZ_2\times X)_{\rm mixed} \longrightarrow \RR/\ZZ
\end{equation}
defines a bordism invariant.

\paragraph{Main claim.}
Using the Smith isomorphism reviewed in section~\ref{sec:smith}, we may equivalently consider $\frac{1}{2} \widetilde{\eta}_E(M,a)$ as a bordism invariant for $\widetilde{\Omega}_2^{\pinm}(X)$. Let
\begin{equation} 
        \Sigma = {\rm PD}(a)\subset M
\end{equation}
be the surface Poincar\'e dual to $a$. As in section~\ref{sec:smith}, the spin structure on
$M$ induces a $\pinm$ structure on $\Sigma$, and the classifying map restricts to $f|_\Sigma : \Sigma \to X$. Our main claim is now that
\begin{equation}
\pm \frac{1}{2} \widetilde{\eta}_E(M,a)=\frac{1}{2} \langle w_2(E|_\Sigma), [\Sigma] \rangle-\frac{1}{4} Q_\Sigma (w_1(E|_\Sigma)) \mod 1\,, \label{eq:eta-to-pin-invariants}
\end{equation}
corresponding to the two bordism invariants explained in \eqref{eq:deg2inv} and \eqref{eq:bordinv2}.
The appearance of Stiefel-Whitney classes in this formula will allow us to connect the anomaly with the topology of the target space in string theory. The $\pm$ sign on the left-hand side is a universal overall sign that we could in principle compute, but which would require us to chase the conventions for the eta invariant more carefully. It will not be needed for our application below.

\paragraph{Proof for line bundles.} We set everything up to prove \eqref{eq:eta-to-pin-invariants}. Let us first show it for line bundles, $E=L_{a'}$ for $a' \in H^1(M,\ZZ_2)$. Then
\begin{align} 
\frac{1}{2} \widetilde{\eta}_{L_{a'}}(M,a)&=\frac{1}{2}\big(\eta(\mathcal{D}_M^{L_{a+a'}})-\eta(\mathcal{D}_M^{L_{a}})-\eta(\mathcal{D}_M^{L_{a'}})+\eta(\mathcal{D}_M^0)\big) \nonumber\\
&\in \mho_3^\spin(B\ZZ_2 \times B\ZZ_2)_\mathrm{mixed} \cong e(\ZZ_2,\ZZ_2)\cong \ZZ_4\,.
\end{align}
The isomorphism to $\ZZ_4$ follows from the explicit description of the extension \eqref{eq:extension group law} and the corresponding bordism invariant is $Q_{\mathrm{PD}(a)}(a')$; see \eqref{eq:bordinv2}. Since both $\frac{1}{2} \widetilde{\eta}_{L_{a'}}(M,a)$ and $Q_{\mathrm{PD}(a)}(a')$ are bordism invariants, it follows that they are proportional. 

The constant of proportionality can be fixed by evaluating both invariants on a generator of $\Omega_3^\spin(B\ZZ_2 \times B\ZZ_2)_\mathrm{mixed}$. Such a generator is given by $\RP^3$ (with any of the two possible spin structure choices) and $a=a' \in H^1(\RP^3,\ZZ_2)$ the unique non-trivial element, so that $L_a=L_{a'}=L$ is the tautological line bundle. The eta invariant on $\RP^3$ can be easily evaluated using the formula of Gilkey \cite{gilkey}. For a spherical space form $S^3/\Gamma$, with $\Gamma\subset\U(2)$ a finite group acting freely on $S^3\subset\mathbb{C}^2$, and the Dirac operator twisted by the flat bundle associated to a virtual representation $\rho$ of $\Gamma$, it reads \cite[Lemma~2.3(a)]{gilkey}
\begin{equation}
\eta(\mathcal{D}^{\rho}_{S^3/\Gamma})=\frac{1}{|\Gamma|}\sum_{1\neq g\in\Gamma}\mathrm{Tr}\,\rho(g)\,\frac{\sqrt{\det g}}{\det(\mathds{1}_2-g)}\,. \label{eq:gilkey}
\end{equation}
Here, $g\mapsto\sqrt{\det g}$ has to be a character of $\Gamma$, and the choice of square root defines the spin structure on $S^3/\Gamma$. For $\RP^3$, we have $\Gamma=\{\pm\mathds{1}_2\}$, so only $g=-\mathds{1}_2$ contributes, with $\mathrm{Tr}\,\rho(g)=(-1)^n$ for the line bundle $L^n$.

We then have
\begin{equation} 
\eta(\mathcal{D}_{\RP^3}^{L^n})=\frac{1}{2} (-1)^n \frac{\sqrt{\det (-\mathds{1}_2)}}{\det (\mathds{1}_2+\mathds{1}_2)}=\pm \frac{1}{8} (-1)^n\,.
\end{equation}
The choice of square root corresponds to the choice of spin structure and gives the $\pm$ sign ambiguity.
For the reduced eta invariant, we therefore obtain
\begin{equation} 
\frac{1}{2} \widetilde{\eta}_{L_{a'}}(\RP^3,a)=\frac{1}{4}\,.
\end{equation}
Similarly, we can evaluate $Q_{\mathrm{PD}(a)}(a')$. We have $\mathrm{PD}(a)=\RP^2$. Since $H^\bullet(\RP^2,\ZZ_2)\cong \ZZ_2[a]/(a^3)$, the intersection pairing is simply the product on the one-dimensional vector space $\ZZ_2  \cong \ZZ/2\ZZ$. There are two possible quadratic refinements, namely $Q_\Sigma(a)=\pm a^2 \in \ZZ/4\ZZ$ corresponding to the two $\pinm$ structures. For the generator, we thus also find $\frac{1}{4}Q_{\mathrm{PD}(a)}(a')=\pm \frac{1}{4}$. Thus \eqref{eq:eta-to-pin-invariants} is verified for line bundles.

\paragraph{Reducing the case of vector bundles to line bundles.} We now discuss the general case. For the most general choice of real vector bundle, $E$ is determined by a classifying map into $B\mathrm{O}(r)$ with $r$ the rank of the vector bundle. Thus we may assume that $X \cong B\mathrm{O}(r)$. In that case, we have
\begin{equation} 
\frac{1}{2} \widetilde{\eta}_E(M,a) \in \mho_3^\spin(B\ZZ_2 \times B\mathrm{O}(r))_\mathrm{mixed} \cong e(H^2(B\mathrm{O}(r),\ZZ_2),H^1(B\mathrm{O}(r),\ZZ_2))\,.
\end{equation}
Since $H^*(B\mathrm{O}(r),\ZZ_2)\cong\ZZ_2[w_1,\dots,w_r]$, $H^1(B\mathrm{O}(r),\ZZ_2)$ is generated by $w_1$, while $H^2(B\mathrm{O}(r),\ZZ_2)$ is spanned by $w_1^2$ and $w_2$ (by $w_1^2$ alone for $r=1$).
No other more subtle topological information about the vector bundle $E$ is needed. In particular, by the splitting principle, the characteristic classes $w_1$ and $w_2$ can be detected on sums of line bundles. The upshot is that it suffices to check \eqref{eq:eta-to-pin-invariants} for sums of line bundles; it then follows for general vector bundles. So let us assume that $E$ splits on $M$, so that
\begin{equation}
       E|_\Sigma \simeq L_1\oplus\cdots\oplus L_r\,,
        \qquad
        u_i:=w_1(L_i)\in H^1(\Sigma,\ZZ_2)\,.
\end{equation}
Using the quadratic refinement property from \eqref{eq:quadraf}, we obtain, in $\ZZ_4$,
\begin{equation}
     \sum_i Q_\Sigma(u_i)
        =
        Q_\Sigma\left(\sum\nolimits_i u_i\right)
        -
        2\sum_{i<j}\left\langle u_i\cup u_j,[\Sigma]\right\rangle\,.
\label{eq:splitting-principle-Q}
\end{equation}
But
\begin{equation}
       w_1(E|_\Sigma)=\sum_i u_i\,,
        \qquad
        w_2(E|_\Sigma)=\sum_{i<j}u_i\cup u_j\,.
\end{equation}
Substituting this into \eqref{eq:splitting-principle-Q} gives exactly \eqref{eq:eta-to-pin-invariants}. This finishes the proof.

\paragraph{The anomaly.}
We can now read off the anomaly as an element of the Pontryagin dual of the
mixed bordism group. Under the identification
\begin{equation} 
        \mho_3^{\spin}(B\ZZ_2\times X)_{\rm mixed}
        \cong
        e\bigl(H^2(X,\ZZ_2),H^1(X,\ZZ_2)\bigr)\,,
\end{equation}
the anomaly of the left-moving Majorana--Weyl fermions valued in $E$ is the
class
\begin{equation}
        \bigl(w_1(E),w_2(E)\bigr)
        \in
        e\bigl(H^2(X,\ZZ_2),H^1(X,\ZZ_2)\bigr)\,.
\label{eq:fermion-anomaly-w1w2}
\end{equation}
This also matches the group law found in section~\ref{sec:extension}. Indeed, if
$E=E_1\oplus E_2$, then the Whitney sum formula gives
\begin{align}
     w_1(E)&=w_1(E_1)+w_1(E_2)\,,\\
     w_2(E)&=w_2(E_1)+w_2(E_2)+w_1(E_1)\cup w_1(E_2)\,.
\end{align}
Therefore, the anomaly classes add according to the extension law \eqref{eq:extension group law}. This gives the desired
identification between the eta invariant anomaly of the worldsheet fermions and
the bordism invariants computed in section~\ref{sec:mixed bordism groups}.

\section{Constraints on target space from worldsheet anomaly cancellation} \label{sec:target space constraints}
After the preparations of sections~\ref{sec:mixed bordism groups} and \ref{sec:eta invariants}, we discuss now the consistency conditions that a GSO projection imposes on the target space geometry. We will assume throughout that the target space geometry is of the form $\RR^{1,1} \times X$, so that we can work in light-cone gauge. We will comment on generalizations that do not assume light-cone gauge in the discussion in section~\ref{sec:discussion}.
\subsection{Smooth target space}\label{sec:smoothts}
We begin by discussing the case of a smooth target space. In light-cone gauge, the worldsheet theory is described by an $\mathcal{N}=(1,1)$ supersymmetric sigma model with the 8-dimensional target space $X$. The action takes the form 
\begin{multline}
S=\frac{1}{4\pi\alpha'} \int_\Sigma \d^2 x \, \big[G_{\mu \nu}(\phi)\partial \phi^\mu \bar{\partial} \phi^\nu +G_{\mu\nu}(\phi) (\psi^\mu_\mathrm{L} \bar{\mathscr{D}} \psi^\nu_\mathrm{L}+ \psi^\mu_\mathrm{R} \mathscr{D} \psi^\nu_\mathrm{R})\\
+\tfrac{1}{2} R_{\mu \nu \rho \sigma} (\phi)\psi^\mu_\mathrm{L}\psi^\nu_\mathrm{L}\psi^\rho_\mathrm{R}\psi^\sigma_\mathrm{R}\big]+\frac{i}{4\pi \alpha'}\int_\Sigma \phi^* B\,. \label{eq:non-linear sigma model}
\end{multline}
Here, $\phi: \Sigma \to X$ describes the bosonic coordinate embedding into target space $X$. The covariant derivatives of the fermions are defined as
\begin{subequations}
\begin{align} 
\bar{\mathscr{D}}\psi_\mathrm{L}^\mu=\bar\partial \psi_\mathrm{L}^\mu+\big(\tensor{\Gamma}{^\mu_{\nu \rho}}(\phi)+\tfrac{1}{2} \tensor{H}{^\mu_{\nu\rho}}(\phi)\big) \bar{\partial} \phi^\nu \psi_\mathrm{L}^\rho\,, \\
\mathscr{D}\psi_\mathrm{R}^\mu=\partial \psi_\mathrm{R}^\mu+\big(\tensor{\Gamma}{^\mu_{\nu \rho}}(\phi)-\tfrac{1}{2} \tensor{H}{^\mu_{\nu\rho}}(\phi)\big) \partial \phi^\nu \psi_\mathrm{R}^\rho\,. 
\end{align}\label{eq:sigma model covariant derivatives}%
\end{subequations}
We follow the conventions of \cite[Section 12.3]{Polchinski:1998rr}. The term involving the $B$-field can be defined in the following standard way. If $\Omega_2^{\mathrm{SO}}(X) \cong H_2(X,\ZZ)=0$, $\phi$ can be extended to a 3-manifold $M$ with $\partial M=\Sigma$. Independence of the chosen extension quantizes $B$. When $H_2(X,\ZZ) \ne 0$, one gets different topological sectors labelled by worldsheet instantons and can associate theta-angles to these sectors labelled by $\Hom(H_2(X,\ZZ),\U(1))$.

\paragraph{Anomalies.} Classically, \eqref{eq:non-linear sigma model} has both the symmetries $(-1)^{\FL}$ acting by $\psi_\mathrm{L} \mapsto -\psi_\mathrm{L}$ and $(-1)^{\FR}$ acting by $\psi_\mathrm{R} \mapsto -\psi_\mathrm{R}$. $(-1)^{\FL}$ also acts on the left-moving $\beta\gamma$ ghosts as $\beta_\mathrm{L} \mapsto -\beta_\mathrm{L}$ and $\gamma_\mathrm{L} \mapsto -\gamma_\mathrm{L}$ and similarly for $(-1)^{\FR}$.

The worldsheet string theory is obtained by gauging these symmetries. Note that
$(-1)^{\Ftot}=(-1)^{\FL}(-1)^{\FR}$ remains non-anomalous at the quantum level, e.g.\ because this symmetry is respected by a massive fermion and thus we can use Pauli--Villars regularization. However, $(-1)^{\FL}$ or $(-1)^{\FR}$ can pick up an anomaly, since no regulator preserves the symmetry. 

For the anomaly, the details of the action \eqref{eq:non-linear sigma model} do not matter. The anomaly originates from the fermion path integral measure and we only require the information that $\psi_{\mathrm{L}}$ and $\psi_{\mathrm{R}}$ are sections of the following bundles,
\begin{equation} 
\psi_\mathrm{L} \in \Gamma(\mathcal{S} \otimes \phi^* TX \otimes L_a)\,, \qquad \psi_\mathrm{R} \in \Gamma(\mathcal{S} \otimes \phi^* TX)\,,
\end{equation}
where $\mathcal{S}$ is the worldsheet spin structure and $L_a$ is the $\ZZ_2$ line bundle over the worldsheet $\Sigma$ associated to $a \in H^1(\Sigma,\ZZ_2)$. The spin structure associated to $\psi_\mathrm{L}$ is then $\mathcal{S} \otimes L_a$. The mixed anomaly is given by
\begin{equation} 
\exp\left(\pi i \big[\eta(\mathcal{D}_{\phi^* TX \otimes L_a})-\eta(\mathcal{D}_{\phi^* TX}) \big]\right)\,. \label{eq:anomaly eta invariant smooth target space}
\end{equation}
This is of the form discussed in section~\ref{subsec:computation anomaly} for $E=\phi^*TX$. Thus, we can apply the main result of section~\ref{subsec:computation anomaly} and determine the anomaly to be \eqref{eq:fermion-anomaly-w1w2}
\begin{equation} 
(w_1(X),w_2(X)) \in \mho_3^\spin(B\ZZ_2 \times X)_\mathrm{mixed} \cong e\big(H^2(X,\ZZ_2), H^1(X,\ZZ_2)\big)\,. \label{eq:anomaly cancellation condition smooth target space}
\end{equation}
The additional purely gravitational terms originating from the passage to the reduced eta invariant \eqref{eq:reduced-eta-E} do not contribute since the rank of $E$ is 8. 
Thus, anomaly cancellation requires the target spacetime to be a spin manifold with $w_1(X)=w_2(X)=0$.


\paragraph{Different GSO projections.}
Assuming that the target space is spin, it might still admit different GSO projections. We now fully classify the worldsheet theta angles appearing and discuss their spacetime interpretation.

These theta angles are classified on the worldsheet by the bordism group $\mho^\spin_2(B\ZZ_2 \times X)$. We computed it in  \eqref{eq:mho 2} and listed the corresponding bordism invariants in \eqref{eq:mho2 bordism invariants}.
These five bordism invariants have a clear geometric meaning.
\begin{itemize}
    \item $\mathrm{Arf}^\mathrm{L}$ and $\mathrm{Arf}^\mathrm{R}$. These are the purely gravitational pieces independent of the target space geometry. They label the different types of string theories. This naively gives four different string theories, although they are pairwise related by target space orientation reversal, which includes the discrete theta angle $\exp(\pi i(\mathrm{Arf}^\mathrm{L}+\mathrm{Arf}^\mathrm{R}))$. As discussed in \cite{Kaidi:2019pzj}, after taking these identifications into account there are only two physically inequivalent GSO projections corresponding to type IIA and type IIB string theory.
    \item $\exp(2\pi i \langle \phi^* b,  [\Sigma] \rangle)$ gives a continuous family of theta angles, together with a discrete-torsion piece coming from the torsion of $H_2(X,\ZZ)$. This is simply the $B$-field term in the bosonic sigma model action (with $b=B$). Thus these theta angles correspond to the $B$-field background and are part of the conformal manifold of the worldsheet theory. These theta angles have nothing to do with the fermions or the GSO projection. 
    \item $\exp(\pi i Q_\Sigma^\mathrm{L}(a^\mathrm{L}))$ and $\exp(\pi i Q_\Sigma^\mathrm{R}(a^\mathrm{R}))$. These are the most interesting. The target space theory has two exact $\ZZ_2$ symmetries that are visible in string perturbation theory \cite{Tachikawa:2018njr}.\footnote{It in fact has a third, corresponding to worldsheet parity reversal, but analyzing it would require us to also include orientifolds into our analysis. We comment on such an extension in section~\ref{sec:discussion}.} The first, $\smash{\widehat{\smash{(-1)^{\FL}}\vphantom{\rule{0pt}{1.8ex}}}}$, is the quantum symmetry that arises after gauging $(-1)^{\FL}$ on the worldsheet, acting as $-1$ on the R-R fields, one of the two dilatini and on one of the two gravitini. One could have equivalently gauged $(-1)^{\FR}$ on the worldsheet, which leads to a physically equivalent theory and a quantum symmetry $\smash{\widehat{\smash{(-1)^{\FR}}\vphantom{\rule{0pt}{1.8ex}}}}$, which acts as $-1$ on the R-R fields and the two other fermions. The second symmetry is spacetime fermion parity $(-1)^{\Fs}=\smash{\widehat{\smash{(-1)^{\FL}}\vphantom{\rule{0pt}{1.8ex}}}}\smash{\widehat{\smash{(-1)^{\FR}}\vphantom{\rule{0pt}{1.8ex}}}}$, which acts as $-1$ on all target space fermions. In terms of the invariants above, $a^\mathrm{L}$ and $a^\mathrm{R}$ are the backgrounds for $(-1)^{\FL}$ and $(-1)^{\FR}$ on the worldsheet; thus $\smash{\widehat{\smash{(-1)^{\FL}}\vphantom{\rule{0pt}{1.8ex}}}}$ or $\smash{\widehat{\smash{(-1)^{\FR}}\vphantom{\rule{0pt}{1.8ex}}}}$ corresponds to a background $a^{\mathrm{L}/\mathrm{R}}$ while $(-1)^{\Fs}$ corresponds to the diagonal background $a^\mathrm{L}+a^\mathrm{R}$, i.e.\ the spin structure shift. We can also describe these backgrounds directly in target space. The $a^\mathrm{L}$ (or $a^\mathrm{R}$) background does not mix in with the spin structure, so in spacetime we obtain a standard spin structure, equipped with a $\ZZ_2$ gauge bundle for $\smash{\widehat{\smash{(-1)^{\FL}}\vphantom{\rule{0pt}{1.8ex}}}}$ (or $\smash{\widehat{\smash{(-1)^{\FR}}\vphantom{\rule{0pt}{1.8ex}}}}$). Other choices of spin structure and $a^\mathrm{L/R}$ background are obtained by tensoring with flat $\ZZ_2$ line bundles for $(-1)^{\Fs}$ and $\smash{\widehat{\smash{(-1)^{\FL}}\vphantom{\rule{0pt}{1.8ex}}}}$ (or $\smash{\widehat{\smash{(-1)^{\FR}}\vphantom{\rule{0pt}{1.8ex}}}}$). This is in one-to-one correspondence with an element of $H^1(X,\ZZ_2) \oplus H^1(X,\ZZ_2)$, which is what appears on the worldsheet. Let us also note that we do not get a canonical isomorphism with such structures in target space. Indeed, the target space data (spin structure and $\ZZ_2$ gauge bundle) form a torsor over $H^1(X,\ZZ_2)\oplus H^1(X,\ZZ_2)$, while the consistent definitions of the GSO-projected worldsheet partition function form a torsor over the group of theta angles, with no canonical reference choice on either side \cite{Witten:2019bou}. The correspondence is therefore between two torsors: the discrete theta angle relating two worldsheet partition functions is matched with the difference, in $H^1(X,\ZZ_2)\oplus H^1(X,\ZZ_2)$, of the corresponding target space data. The spin structure part was already discussed in different language in \cite{Atick:1988si, Acharya:2020hsc}. 
\end{itemize}
Our computation shows that this exhausts the possible GSO projections; there are no exotic ones.

\subsection{Flat space orbifolds}\label{sec:flatspaceorb}
The next class of target spaces that we are going to consider are flat space orbifolds, where the expectation from supergravity is less clear. Consistency of toroidal orbifolds from the perspective of global anomaly cancellation on the worldsheet was studied in explicit examples using the mapping torus construction in \cite{Freed:1987qk}. The case of orbifolds by cyclic groups was later studied under the lens of Dai--Freed anomaly cancellation, using bordism groups in \cite{Cheng:2026abk}. We now formulate a general criterion for the cancellation of worldsheet Dai--Freed anomalies in flat-space orbifolds of type II strings.

\paragraph{Fermion anomalies.} With our preparations, the anomaly is straightforward to determine. Let $G$ be the orbifold group. $G$ acts on the target space by a representation $\rho:G 
\longrightarrow \mathrm{O}(8)$. Then the vector bundle $E$ of the fermions is a $G$-bundle and therefore classified by a map into $BG$. $\rho$ induces a map $B\rho:BG \longrightarrow B\mathrm{O}(8)$. We have by definition $w_k(E)=(B\rho)^*w_k \in H^k(BG,\ZZ_2)$, the Stiefel--Whitney classes of the representation $\rho$.
From \eqref{eq:fermion-anomaly-w1w2}, it follows immediately that the anomaly of the GSO projection is characterized by
\begin{equation} 
(w_1(\rho),w_2(\rho)) \in \mho_3^\spin(B\ZZ_2 \times BG)_\mathrm{mixed} \cong e\big(H^2(BG,\ZZ_2) ,H^1(BG,\ZZ_2)\big)\,. \label{eq:anomaly-free field orbifold}
\end{equation}
Here $w_1(\rho)$ measures the obstruction of the representation $\rho$ to lift to $\mathrm{SO}(8)$. It has the explicit formula $w_1(\rho)=\det \rho: G \longrightarrow \ZZ_2$, which can be thought of as an element of $H^1(G,\ZZ_2)$. $w_2(\rho)$ measures the obstruction of the representation $\rho$ to lift to $\Spin(8)$. Thus, the GSO projection on the orbifold CFT is consistent provided that $\rho$ lifts to a Spin representation.

\paragraph{Representation theoretic view.} The necessity of the previous criterion can also be seen more elementarily. The GSO projection and the $G$-orbifold together are a $\ZZ_2 \times G$ orbifold. We can try to first perform the $\ZZ_2$-orbifold and then the $G$-orbifold. A necessary condition for the consistency of the orbifold is that the flat space GSO-projected Hilbert space carries a proper $G$-action. Consider the ground states of the R-NS sector. Famously, they transform in the $\Spin(8)$ representation $\mathbf{8}_\mathbf{s} \otimes \mathbf{8}_\mathbf{v}$ (or $\mathbf{8}_\mathbf{c} \otimes \mathbf{8}_\mathbf{v}$). To define the $G$-action, it is thus necessary to uplift the representation $\rho$ to a spinor representation, which leads to the condition of the vanishing of the first two Stiefel--Whitney classes of the representation as in \eqref{eq:anomaly-free field orbifold}.

\paragraph{Different GSO projections and discrete torsion.} We now turn to the theta angles classified by the bordism group
$\mho^\spin_2(B\ZZ_2 \times BG)$. We computed this group in \eqref{eq:mho 2}, and listed the corresponding bordism invariants in \eqref{eq:mho2 bordism invariants}. The purely gravitational invariants $\mathrm{Arf}^{\mathrm{L}/\mathrm{R}}$ have the same interpretation as for smooth target spaces: they distinguish the inequivalent GSO projections giving the type IIA and type IIB strings. The invariants $ \exp(\pi i Q^{\mathrm{L}/\mathrm{R}}_\Sigma(\phi^* a^{\mathrm{L}/\mathrm{R}}))$ with $a^{\mathrm{L}/\mathrm{R}}\in H^1(BG,\ZZ_2)$ should be viewed as the orbifold version of the remaining discrete target space choices. For a flat-space orbifold with representation $ \rho\to \mathrm{SO}(8)$, these choices correspond, when they exist, to the different lifts of $\rho$ to $\Spin(8)$, which can be chosen independently for left- and right-movers. Finally, the invariant $ \exp (2\pi i\big\langle \phi^* b [\Sigma]\big\rangle)$ with $b\in H^2(BG,\U(1))$, is the usual discrete torsion phase of the orbifold \cite{Vafa:1986wx,Narain:1986qm,Narain:1990mw}.

\subsection{General orbifolds} \label{subsec:orbifold target spaces}
We now combine the previous two cases and consider a general sigma model with an orbifold target space $X=\hat{X}/G$. For flat-space orbifolds, the anomalies were associated to the linear action of $G$ on the eight free fermions. For a general orbifold, this is no longer enough: the fermions couple to the full equivariant tangent bundle, which also remembers the topology of  $\hat{X}$, the variation of the group action over $\hat{X}$, and how the local actions near fixed loci fit together globally. Thus, we now derive a constraint for general orbifolds to be free of such anomalies, and show how it reduces to the constraints derived previously in sections \ref{sec:smoothts} and \ref{sec:flatspaceorb}.

\paragraph{Fermion vector bundle.} 
The worldsheet path integral now involves a sum over flat $G$-bundles on $\Sigma$, each specified up to conjugation by a holonomy homomorphism  $\varrho:\pi_1(\Sigma)\to G$. A choice of $\varrho$ fixes a twisted sector. In such a sector, the worldsheet embedding map $\Sigma$ to $\hat{X}$ is not, in general, single-valued; instead, after lifting to the universal cover $\widetilde{\Sigma}$, it becomes become an ordinary map to $\hat{X}$ whose monodromy is prescribed by the orbifold group action. More precisely, the embedding $\phi$ is a map from the universal cover $\widetilde{\Sigma}$ to $\hat{X}$ obeying the twisted boundary condition $\phi(\gamma\cdot z)=\varrho(\gamma)\cdot\phi(z)$ for $\gamma\in\pi_1(\Sigma)$, where on the right $G$ acts through the orbifold action on $\hat{X}$. The pair $(\phi,\varrho)$ is the same data as a single map $\Phi:\Sigma\to(\hat{X}\times EG)/G$ into the Borel construction: $\varrho$ supplies the underlying map to $BG=EG/G$, and $\phi$ supplies the lift along the $\hat{X}$-fibre. The worldsheet fermions then take values in the pulled-back equivariant tangent bundle, $E=\Phi^*\big((T\hat{X}\times EG)/G\big)$.

Another point of view of this is to think about the target space as a stack $[\hat{X}/G]$, which retains the information of the $G$-action \cite{Sharpe:2001bs}. The space of maps into $[\hat{X}/G]$ can be identified with maps into $(\hat{X} \times EG)/G$.

\paragraph{Anomaly.} In any case, this discussion shows that the anomaly lives in
\begin{equation} 
\mho_{3}^\spin(B\ZZ_2 \times (\hat{X} \times EG)/G)_\mathrm{mixed} \cong e\big(H_G^2(\hat{X},\ZZ_2) , H_G^1(\hat{X},\ZZ_2)\big)\,,
\end{equation}
where $H_G^k(\hat{X},\ZZ_2):=H^k((\hat{X} \times EG)/G,\ZZ_2)$ denotes equivariant cohomology. From the same discussion as in the previous two cases, the existence of a GSO projection imposes
\begin{equation} 
w_1^G(\hat{X})=w_2^G(\hat{X})=0\,, \label{eq:anomaly cancellation condition orbifold target space}
\end{equation}
i.e.\ the first two equivariant Stiefel--Whitney classes of $\hat{X}$ have to vanish. In other words, $\hat{X}$ has to admit a $G$-equivariant spin structure, see e.g.\ \cite{Atiyah1970}. 

The Borel construction admits the fibration
\begin{equation} 
\hat{X} \longrightarrow (\hat{X} \times EG)/G \longrightarrow BG\,.
\end{equation}
In the two cases discussed above, one of these terms was trivial: $BG\sim \mathrm{pt}$ for a smooth target space and $\hat{X} \sim \mathrm{pt}$ for a flat space orbifold. In both these cases, the fibration is trivial, and we recover the previous condition.
\paragraph{Free group actions.} As another consistency check, we can consider the case when $G$ acts freely on $\hat{X}$, in which case the orbifold coincides with the sigma model on $X=\hat{X}/G$. $(\hat{X} \times EG)/G \sim \hat{X}/G=X$ is a homotopy equivalence since $(\hat{X} \times EG)/G \to \hat{X}/G$ is a fiber bundle with contractible fiber $EG$. Thus, equivariant cohomology on $\hat{X}$ can naturally be identified with ordinary cohomology on $X$. Under the homotopy equivalence, the equivariant tangent bundle $(T\hat{X} \times EG)/G$ is identified with the tangent bundle $T(\hat{X}/G)$. Therefore, the equivariant Stiefel--Whitney classes on $\hat{X}$ are identified with the Stiefel--Whitney classes on the quotient, $w_k^G(\hat{X})=w_k(\hat{X}/G)$. This reduces the anomaly cancellation to the smooth target space condition \eqref{eq:anomaly cancellation condition smooth target space}.

\paragraph{Orbifold locus.}
From $w_1^G(\hat{X})=0$ it follows that $G$ acts by orientation-preserving
diffeomorphisms. We now show that the orbifold locus of the $G$–action has even
codimension in $\hat{X}$. In particular, end-of-the-world branes consistent with the GSO projection cannot be constructed in string perturbation theory in this fashion.

Indeed, if $x$ lies in the orbifold locus, then its stabilizer
$H=\mathrm{Stab}(x)$ is non-trivial and acts on the normal space
$\mathcal{N}\cong\RR^8$ by linear maps in $\mathrm{SO}(8)$.
A neighborhood of $x$ in the quotient is modeled on $\mathcal{N}/H$, and the
orbifold locus is locally the image of
$\bigcup_{1\neq h\in H}\mathcal{N}^h$, where
$\mathcal{N}^h=\ker(h-\mathrm{id})$ is the fixed-point set of $h$.
The codimension of $\mathcal{N}^h$ in $\mathcal{N}$ is the number of
eigenvalues of $h$ different from $1$. These eigenvalues lie on the unit
circle and occur either in complex conjugate pairs or as the real eigenvalue
$-1$. Since $h\in\mathrm{SO}(8)$ has determinant $1$, the eigenvalue $-1$
appears with even multiplicity. Hence $\dim\mathcal{N}-\dim\mathcal{N}^h$ is
even for every $1\neq h\in H$, and therefore each stratum of the orbifold locus has even codimension.

\paragraph{Example.} Let us work out an example to see how these anomalies are manifested in orbifolds that are not equivariantly spin. An instructive case is that of the orbifold $\TT^2/\ZZ_n \times \TT^2$. We take $\hat{X}=\RR^4\times\TT^2\times\TT^2$ and $G=\ZZ_n$, so that $X$ is $\RR^4$ times this orbifold.
We will focus on the following four quotients with group $\ZZ_n$ with $n=2,\,3,\,4,\,6$. They act on the first torus by rotations with angle $\frac{2\pi}{n}$ and trivially on the second torus.\footnote{One may in addition let $\ZZ_n$ act on the second torus by a shift of $\frac{1}{n}$ along a lattice vector. The action is then free, and the quotient is a smooth hyperelliptic surface. It is spin for all $n$, being diffeomorphic to a circle times the mapping torus of the rotation, and hence parallelizable. Apart from making the quotient smooth, the shift does not enter the computation below, with one caveat: for even $n$, it has to be accompanied by spacetime fermion parity $(-1)^{\Fs}$, which modifies the periodicity of $\zeta$ in $r$ assumed below. We thank Juan Ospina and the anonymous referee for corrections on this example.}
The case of $n=4$ requires the first torus to stem from a square lattice, while $n=3$ and $6$ require a hexagonal lattice. For even $n$, any lift $\tilde{R}\in\Spin(4)$ of the rotation satisfies $\tilde{R}^n=-1$, so that the $\ZZ_n$ action does not lift to spinors, i.e.\ $\hat{X}$ admits no $\ZZ_n$-equivariant spin structure.

The one-loop partition function of these models is straightforward to work out. Let us define
\begin{equation} 
\vartheta_{\svecsub{\alpha}{\beta}}(\tau)=\mathrm{e}^{\pi i \beta^2 \tau} \vartheta_1(\alpha+\beta \tau,\tau)\,,
\end{equation}
which behaves nicely under modular transformations. Indeed, it satisfies
\begin{subequations}
\begin{align} 
\vartheta_{\svecsub{\alpha}{\beta}}(-\tfrac{1}{\tau})&=-i\sqrt{-i \tau}\mathrm{e}^{-2\pi i \alpha\beta} \vartheta_{\svecsub{-\beta}{\alpha}}(\tau)\,, \\
\vartheta_{\svecsub{\alpha}{\beta}}(\tau+1)&=\mathrm{e}^{\pi i (\beta^2+\frac{1}{4})} \vartheta_{\svecsub{\alpha+\beta}{\beta}}(\tau)\,, \\
\vartheta_{\svecsub{\alpha+1}{\beta}}(\tau)&=-\vartheta_{\svecsub{\alpha}{\beta}}(\tau)\,, \label{eq:theta alpha+1}\\
\vartheta_{\svecsub{\alpha}{\beta+1}}(\tau)&=-\mathrm{e}^{-2\pi i \alpha} \, \vartheta_{\svecsub{\alpha}{\beta}}(\tau)\,. \label{eq:theta beta+1}%
\end{align}
\end{subequations}
The partition function takes the form
\begin{equation} 
Z=\frac{1}{|\eta^4|^2}\sum_{\alpha,\beta,\tilde{\alpha},\tilde{\beta}\in \{0,1/2\}} \sum_{r,s \in \{0,1/n,\dots,(n-1)/n\}} \zeta\big(\svec{\alpha}{\beta}, \svec{\tilde{\alpha}}{\tilde{\beta}}, \svec{r}{s}\big)Z^\mathrm{bos}_{\svecsub{r}{s}}\vartheta_{\svecsub{\alpha}{\beta}}^3 \vartheta_{\svecsub{\alpha+r}{\beta+s}}\bar{\vartheta}_{\svecsub{\tilde{\alpha}}{\tilde{\beta}}}^3\bar{\vartheta}_{\svecsub{\tilde{\alpha}+r}{\tilde{\beta}+s}}\,.
\end{equation}
Here, the sum over $\alpha$ and $\beta$ ($\tilde{\alpha}$ and $\tilde{\beta}$) performs the sum over left-moving (right-moving) spin structures, while $r$ and $s$ parametrize the twisted sectors of the $\ZZ_n$ orbifold. Note that $\eta$ denotes the Dedekind eta function (not the eta invariant). We will not need the precise value of the bosonic twisted partition function. $\zeta$ parametrizes arbitrary phases of the twisted sectors. The consistency of the theory boils down to whether there is a consistent assignment of phases for the different twisted sectors.\footnote{The conditions from one-loop modular invariance are necessary but in general not sufficient: a consistent orbifold must also have well-defined higher-genus amplitudes. The first additional constraint arises at genus two, where factorization of the partition function ties the higher-genus phases to the one-loop data and can further restrict the allowed discrete torsion \cite{Vafa:1986wx}. }
It is convenient to extend them outside of the range of summation via \eqref{eq:theta alpha+1} and \eqref{eq:theta beta+1} as follows
\begin{subequations}
\begin{align}
\zeta\big(\svec{\alpha}{\beta +1}, \svec{\tilde{\alpha}}{\tilde{\beta}}, \svec{r}{s}\big)&=\mathrm{e}^{2\pi i r}\zeta\big(\svec{\alpha}{\beta}, \svec{\tilde{\alpha}}{\tilde{\beta}}, \svec{r}{s}\big)\,, \label{eq:zeta periodicity 1}\\
\zeta\big(\svec{\alpha}{\beta}, \svec{\tilde{\alpha}}{\tilde{\beta}+1}, \svec{r}{s}\big)&=\mathrm{e}^{-2\pi i r}\zeta\big(\svec{\alpha}{\beta}, \svec{\tilde{\alpha}}{\tilde{\beta}}, \svec{r}{s}\big)\,, \label{eq:zeta periodicity 2}\\
\zeta\big(\svec{\alpha}{\beta}, \svec{\tilde{\alpha}}{\tilde{\beta}}, \svec{r}{s+1}\big)&=\mathrm{e}^{2\pi i (\alpha-\tilde{\alpha})}\zeta\big(\svec{\alpha}{\beta}, \svec{\tilde{\alpha}}{\tilde{\beta}}, \svec{r}{s}\big)\,, \label{eq:zeta periodicity 3}%
\end{align}
\end{subequations}
while $\zeta$ is invariant under shifting $\alpha$, $\tilde{\alpha}$ or $r$ by integers.
Invariance under $\tau \to -\frac{1}{\tau}$ and $\tau \to \tau+1$ implies
\begin{subequations}
\begin{align} 
\zeta\big(\svec{\alpha}{\beta}, \svec{\tilde{\alpha}}{\tilde{\beta}}, \svec{r}{s}\big)\mathrm{e}^{4\pi i(\beta^2-\tilde{\beta}^2)+2\pi i (\beta-\tilde{\beta})s}&=\zeta\big(\svec{\alpha+\beta}{\beta}, \svec{\tilde{\alpha}+\tilde{\beta}}{\tilde{\beta}}, \svec{r+s}{s}\big)\,, \label{eq:zeta T invariance}\\
\zeta\big(\svec{\alpha}{\beta}, \svec{\tilde{\alpha}}{\tilde{\beta}}, \svec{r}{s}\big)\mathrm{e}^{-2\pi i (\alpha-\tilde{\alpha})s-2\pi i (\beta-\tilde{\beta})r} &=\zeta\big(\svec{-\beta}{\alpha}, \svec{-\tilde{\beta}}{\tilde{\alpha}}, \svec{- s}{r}\big)\,.\label{eq:zeta S invariance}%
\end{align} 
\end{subequations}
The anomaly we are after is mixed between the $\ZZ_2$ and $\ZZ_n$ orbifold groups. $\zeta$ is unambiguous in the untwisted sector of the $\ZZ_2$ orbifold or the untwisted sector of the $\ZZ_n$ orbifold,
\begin{equation} 
\zeta\big(\svec{\alpha}{\beta}, \svec{\alpha}{\beta}, \svec{r}{s}\big)=1\,, \quad \zeta\big(\svec{\alpha}{\beta}, \svec{\tilde{\alpha}}{\tilde{\beta}}, \svec{0}{0}\big)=\mathrm{e}^{4\pi i (\alpha \beta+\tilde{\alpha}\tilde{\beta})}\,.
\end{equation}
The latter signs correspond to the standard signs appearing in the flat space GSO projection. For odd $n$, one can easily confirm that the following is a solution
\begin{equation} 
\zeta\big(\svec{\alpha}{\beta}, \svec{\tilde{\alpha}}{\tilde{\beta}}, \svec{r}{s}\big)=\mathrm{e}^{2\pi i n s(\alpha-\tilde{\alpha})+2\pi i(n+1)r(\beta-\tilde{\beta})+4\pi i(\alpha \beta+\tilde{\alpha}\tilde{\beta})}\,.
\end{equation}
Oddness of $n$ is only required for the quasiperiodicity property \eqref{eq:zeta periodicity 3}. 

For even $n$, the equation \eqref{eq:zeta T invariance} quickly leads to a contradiction by applying it twice:
\begin{multline}
\zeta\big(\svec{\alpha}{1/2},\svec{\tilde{\alpha}}{0}, \svec{r}{1/2}\big)=i\, \zeta\big(\svec{\alpha+1/2}{1/2},\svec{\tilde{\alpha}}{0}, \svec{r+1/2}{1/2}\big)\\
=i^2\, \zeta\big(\svec{\alpha+1}{1/2},\svec{\tilde{\alpha}}{0}, \svec{r+1}{1/2}\big)=-\zeta\big(\svec{\alpha}{1/2},\svec{\tilde{\alpha}}{0}, \svec{r}{1/2}\big)\,.
\end{multline}
In the last step, we used the periodicity of $\zeta$ in $\alpha$ and $r$. Thus, there is no consistent orbifold, reflecting the fact that the $\ZZ_n$ action does not lift to spinors. This is in agreement with \eqref{eq:anomaly cancellation condition orbifold target space}. Indeed, restricted to a fixed point of the rotation, $w_2^G(\hat{X})$ reduces to $w_2(\rho)$ for the representation $\rho:\ZZ_n \to \mathrm{O}(4)\subset \mathrm{O}(8)$ given by the rotation, as in section~\ref{sec:flatspaceorb}, and $\rho$ admits a lift to $\Spin(4)$ only when $n$ is odd. This is precisely the case in which the partition function can be made modular invariant.

\section{Discussion} \label{sec:discussion}

In this paper, we analyzed the topological obstruction to performing the type II GSO projection in various target space backgrounds. These obstructions are global anomalies of the worldsheet fermions, detected by spin bordism groups. We determined the relevant mixed bordism group for a simultaneous coupling to $(-1)^{\FL}$ and to target space background data, and identified the corresponding Dai--Freed anomaly in terms of the Stiefel--Whitney classes of the real vector bundle in which the worldsheet fermions take values.

We then applied this criterion to several classes of backgrounds. For smooth target spaces, anomaly cancellation requires the target to be orientable and spin. For flat-space orbifolds, the same condition becomes the requirement that the orbifold representation lift to a spin representation. For a general orbifold target $[\hat{X}/G]$, the anomaly is controlled by the equivariant tangent bundle, and the condition is
\begin{equation}
w_1^G(\hat{X})=w_2^G(\hat{X})=0\,.
\end{equation}
Equivalently, the orbifold must admit a $G$-equivariant spin structure. This formulation incorporates not only the topology of $\hat{X}$ and the linear action of $G$, but also the way the group action varies over $\hat{X}$, including the local representations along orbifold strata. In particular, it reduces to the ordinary spin condition for smooth quotients by free actions, and to the spin-lift condition for flat-space orbifolds.

We also computed the relevant degree-two bordism groups, which label the possible discrete theta angles on the worldsheet. In geometric backgrounds, these reproduce the expected choices of spacetime spin structure and quantum symmetry gauge bundle, together with the usual distinction between the type IIA and type IIB projections and the independent $B$-field theta angles. Thus, within the class of backgrounds considered here, the bordism analysis accounts for the possible GSO projections and does not reveal additional exotic choices.
\bigskip

We now comment on possible extensions and future directions. 

\paragraph{Covariant formalism.} There are several directions in which our results could be extended. First, we assumed a target space of the form $\RR^{1,1} \times X$, so that light-cone gauge can be performed. It is, in principle, better to use a covariant formulation, but this requires proper treatment of the ghosts, as well as dealing with the non-unitarity of the resulting matter theory. The ghosts are non-unitary and require one to properly deal with operators of different picture numbers. In particular, they are not guaranteed to follow the standard anomaly classification. If we naively go ahead and ignore those issues, an explicit calculation, unsurprisingly, shows that they carry the anomaly of $-2$ fermions (as they essentially cancel two fermions). This means that a general 10-dimensional target space $X$ can be incorporated into our analysis by replacing $\phi^*TX$ with the virtual bundle $TX - \varepsilon^{ \oplus 2}$, where $\varepsilon^{\oplus 2}$ is the trivial bundle of rank 2. This does not change anything about the analysis in section~\ref{sec:target space constraints}, and we still conclude that the target space has to carry a spin structure for consistency.

\paragraph{Orientifolds.} Type IIA string theory can also be constructed on non-orientable target spaces. However, such non-orientable target spaces also require non-orientable worldsheets, which are not captured by our analysis. More concretely, for $X$ a non-orientable target space, we can consider $\widetilde{X}$ its orientation double cover, which is equipped with an orientation-reversing $\ZZ_2$ symmetry. The worldsheet theory can then be formulated as an orientifold of the worldsheet sigma model into the target space $\widetilde{X}$, which exchanges left- and right-movers. The relevant worldsheet structure for a flat target space was discussed in \cite{Kaidi:2019tyf}, where it was called a $\mathrm{DPin}$ structure. There should be a similar story generalizing our analysis to the non-orientable target space situation, which should let one discover that type IIA string theory also makes sense on non-orientable manifolds; see \cite{Diaconescu:2000wy, Distler:2009ri, Distler:2010an, Sati:2011rw} for discussions from the target space perspective.

\paragraph{D-branes.} In this paper, we exclusively dealt with closed strings, for which the worldsheet theory is described by a CFT. A natural next step is to consider open strings, whose worldsheet description is given by boundary conformal field theories (BCFTs). In this framework, conformally invariant boundary conditions correspond to D-branes. It has been understood that the presence of boundaries leads to additional consistency conditions associated with anomalies; see, for instance, \cite{Jensen:2017eof}. At the same time, boundaries allow for a richer structure, as the bulk theory need not be anomaly-free on its own, provided that its anomaly is cancelled by suitable boundary degrees of freedom. A seminal example is the Freed–Witten anomaly, where the consistency of the open-string worldsheet path integral imposes nontrivial topological constraints on D-brane worldvolumes and their gauge field data \cite{Freed:1999vc}. In modern language, this can be understood as a relative anomaly cancellation condition involving both bulk and boundary contributions. 

It would be interesting to revisit such questions from the bordism perspective developed here and investigate whether the mixed GSO anomalies considered in this work admit a similar interpretation in the presence of boundaries. In ten-dimensional flat space, such an analysis leads to the familiar K-theoretic classification of D-brane charges \cite{Kaidi:2019tyf}, including both supersymmetric and non-supersymmetric D-branes; see \cite{Witten:2023snr} for a recent review. In orbifold backgrounds, one may expect the analogous analysis to lead to a classification of D-brane charges by $G$-equivariant K-theories, perhaps along the lines of \cite{Braun:2002qa} for the case of type II orientifolds.

\paragraph{Including R-R flux.} The discussion in this paper exclusively considered backgrounds with NS-NS flux. Including R-R flux is difficult from a worldsheet perspective since no simple covariant formalism exists that can formulate string perturbation theory on general backgrounds. However, it might be that for the purposes of analyzing anomalies, not the full worldsheet machinery is needed since the anomaly depends only on relatively coarse data. 

\paragraph{Beyond perturbation theory.} We analyzed consistency within string perturbation theory. If we move beyond perturbation theory, we can uplift type IIB string theory to F-theory and type IIA string theory to M-theory. After uplifting, topological obstructions can become weaker. For example, type IIB string theory does make sense on some non-spin manifolds if one turns on a non-trivial background for the duality group \cite{Cheng:2023owv,Debray:2023yrs}, since one can twist the spin structure by the duality group. Such twists are invisible in string perturbation theory, and it is not possible to formulate string perturbation theory in such a background. 

\paragraph{Non-geometric backgrounds.} The advantage of the worldsheet description is that non-geometric backgrounds can be incorporated without much trouble. We already took a step in that direction by analyzing orbifold target spaces in section~\ref{subsec:orbifold target spaces}. In that case, we of course cannot formulate the consistency conditions in terms of geometric conditions on the target space. 
It is often said that a type II string background is determined by an $\mathcal{N}=(1,1)$ SCFT with $c=\tilde{c}=15$ (setting aside questions about unitarity). However, this does not guarantee the existence of a non-anomalous GSO projection. We instead have to ask for the existence of a non-anomalous
\begin{equation} 
(\ZZ_2 \rtimes \mathcal{N}=1\text{ Vir}) \times (\ZZ_2 \rtimes \mathcal{N}=1\text{ Vir}) \label{eq:worldsheet extended structure}
\end{equation}
structure with $c=\tilde{c}=15$ and with the two $\ZZ_2$'s corresponding to $(-1)^{\FL}$ and $(-1)^{\FR}$ so that the diagonal $\ZZ_2$ is $(-1)^{\Ftot}$.
It is not clear to us whether there is a more useful characterization of such SCFTs.

We should also remark that the structure \eqref{eq:worldsheet extended structure} is guaranteed if we have an $\mathcal{N}=(2,2)$ supersymmetric SCFT with integer-quantized $\mathrm{U}(1)_\mathrm{R}$ charges, as we may take $\FL= J_0 \bmod 2$ and $\FR= \tilde{J}_0 \bmod 2$ in that case where $J_0$ and $\tilde{J}_0$ are the left- and right-moving $\U(1)_\mathrm{R}$ charges. $\mathcal{N}=(2,2)$ supersymmetry with integer-quantized $\mathrm{U}(1)_\mathrm{R}$ charges is needed in order to obtain spacetime supersymmetric backgrounds \cite{Banks:1987cy}. In fact, such models often turn out to describe geometric backgrounds in special limits of moduli space, such as for Gepner models \cite{Gepner:1987qi, Gepner:1987vz}. 

\paragraph{Tachyons and stability.} Our analysis is purely topological and does not rely on spacetime supersymmetry. Accordingly, many of the backgrounds allowed by the anomaly cancellation conditions above are expected to be non-supersymmetric, and may well contain tachyons (at least in some region of scalar field space). This can already happen for smooth targets obtained as freely acting orbifolds: even though there are no fixed points and the quotient is a smooth spin manifold, regions where some cycles become string scale can support winding tachyons. It would be interesting to understand systematically when such instabilities occur, but this question is no longer controlled by the topological data studied here. Rather, it depends on the dynamics of the corresponding string background, including the scalar potentials generated in the absence of supersymmetry. Nevertheless, these dynamical questions are extremely important: one would like to understand the structure of the scalar potentials generated in the absence of supersymmetry, delineate the regions in which tachyons can arise, as was done in e.g. \cite{Fraiman:2023cpa} for heterotic strings, and determine whether the tachyons themselves can be understood along the lines of \cite{Adams:2001sv,Adams:2005rb}. 

\acknowledgments
We thank Bobby Samir Acharya, Joe Davighi, Markus Dierigl, Troels Harmark, Jonathan Heckman, Matthew Heydeman, Bob Knighton, Julio Parra-Martinez, Miguel Montero, Sanjay Raman, Bogdan Stefa\'nski, Ethan Torres, Joaquin Turiaci, Angel M. Uranga, and Cumrun Vafa for helpful discussions. The work of MD is supported in part by a grant from the Simons Foundation (602883,CV) and the DellaPietra Foundation. LE is supported by the European Research Council (ERC) under the European Union’s Horizon 2020 research and innovation programme (grant agreement No 101115511). This paper employed AI tools for verification and language editing.

\bibliographystyle{JHEP}
\inputencoding{latin2}
\bibliography{references}

@article{Freed:1999vc,
    author = "Freed, Daniel S. and Witten, Edward",
    title = "{Anomalies in string theory with D-branes}",
    eprint = "hep-th/9907189",
    archivePrefix = "arXiv",
    doi = "10.4310/AJM.1999.v3.n4.a6",
    journal = "Asian J. Math.",
    volume = "3",
    pages = "819--852",
    year = "1999"
}

@article{Dai:1994kq,
    author = "Dai, Xian-zhe and Freed, Daniel S.",
    title = "{eta invariants and determinant lines}",
    eprint = "hep-th/9405012",
    archivePrefix = "arXiv",
    doi = "10.1063/1.530747",
    journal = "J. Math. Phys.",
    volume = "35",
    pages = "5155--5194",
    year = "1994",
    note = "[Erratum: J.Math.Phys. 42, 2343--2344 (2001)]"
}

@article{Garcia-Etxebarria:2018ajm,
    author = "Garc{\'\i}a-Etxebarria, I{\~n}aki and Montero, Miguel",
    title = "{Dai-Freed anomalies in particle physics}",
    eprint = "1808.00009",
    archivePrefix = "arXiv",
    primaryClass = "hep-th",
    reportNumber = "MPP-2018-188",
    doi = "10.1007/JHEP08(2019)003",
    journal = "JHEP",
    volume = "08",
    pages = "003",
    year = "2019"
}

@mastersthesis{hertl,
    author ="Hertl, Thorsten",
    title ="Pin bordism in low degrees of classifying spaces" ,
    school ="University of G{\"o}ttingen",
    year = 2017,
    url={https://thorsten-hertl.github.io/Research/Masterarbeit.pdf}
}

@article{Cheng:2026abk,
    author = "Cheng, Peng and De Freitas, Hector Parra",
    title = "{Dai-Freed anomalies and level matching in heterotic asymmetric orbifolds}",
    eprint = "2604.19634",
    archivePrefix = "arXiv",
    primaryClass = "hep-th",
    month = "4",
    year = "2026"
}

@article{Thom1954QuelquesPG,
  title={Quelques propri{\'e}t{\'e}s globales des vari{\'e}t{\'e}s diff{\'e}rentiables},
  author={Ren{\'e} Thom},
  journal={Comment. Math. Helv.},
  year={1954},
  volume={28},
  pages={17--86},
  doi={10.1007/BF02566923}
}

@article{Yonekura:2022reu,
    author = "Yonekura, Kazuya",
    title = "{Heterotic global anomalies and torsion Witten index}",
    eprint = "2207.13858",
    archivePrefix = "arXiv",
    primaryClass = "hep-th",
    reportNumber = "TU-1163",
    doi = "10.1007/JHEP10(2022)114",
    journal = "JHEP",
    volume = "10",
    pages = "114",
    year = "2022"
}

@article{Narain:1990mw,
    author = "Narain, K. S. and Sarmadi, M. H. and Vafa, C.",
    title = "{Asymmetric orbifolds: Path integral and operator formulations}",
    reportNumber = "CERN-TH-5846-90, HUTP-90-A053",
    doi = "10.1016/0550-3213(91)90145-N",
    journal = "Nucl. Phys. B",
    volume = "356",
    pages = "163--207",
    year = "1991"
}

@article{Narain:1986qm,
    author = "Narain, K. S. and Sarmadi, M. H. and Vafa, C.",
    title = "{Asymmetric Orbifolds}",
    reportNumber = "HUTP-86-A089",
    doi = "10.1016/0550-3213(87)90228-8",
    journal = "Nucl. Phys. B",
    volume = "288",
    pages = "551",
    year = "1987"
}

@inproceedings{Braun:2002qa,
    author = "Braun, Volker and Stefanski, Jr., B.",
    title = "{Orientifolds and K theory}",
    booktitle = "{NATO Advanced Study Institute and EC Summer School on Progress in String, Field and Particle Theory}",
    eprint = "hep-th/0206158",
    archivePrefix = "arXiv",
    reportNumber = "HU-EP-02-24, AEI-2002-042",
    pages = "369--372",
    month = "6",
    year = "2002"
}

@article{cobordismpin,
 ISSN = {00029939, 10886826},
 doi = {10.1090/S0002-9939-1973-0321123-5},
 URL = {http://www.jstor.org/stable/2039653},
 author = {V. Giambalvo},
 journal = {Proc. Amer. Math. Soc.},
 number = {2},
 pages = {395--401},
 publisher = {American Mathematical Society},
 title = {$\operatorname{Pin}$ and $\operatorname{Pin}'$ {C}obordism},
 urldate = {2026-06-12},
 volume = {39},
 year = {1973}
}

@article{Vafa:1986wx,
    author = "Vafa, Cumrun",
    title = "{Modular Invariance and Discrete Torsion on Orbifolds}",
    reportNumber = "HUTP-86/A011",
    doi = "10.1016/0550-3213(86)90379-2",
    journal = "Nucl. Phys. B",
    volume = "273",
    pages = "592--606",
    year = "1986"
}

@article{Freed:1987qk,
    author = "Freed, Daniel S. and Vafa, Cumrun",
    title = "{Global Anomalies on Orbifolds}",
    reportNumber = "HUTP-86/A090",
    doi = "10.1007/BF01212418",
    journal = "Commun. Math. Phys.",
    volume = "110",
    pages = "349",
    year = "1987",
    note = "[Addendum: Commun.Math.Phys. 117, 349 (1988)]"
}

@article{Seiberg:1986by,
    author = "Seiberg, N. and Witten, Edward",
    title = "{Spin Structures in String Theory}",
    reportNumber = "Print-86-0218 (PRINCETON)",
    doi = "10.1016/0550-3213(86)90297-X",
    journal = "Nucl. Phys. B",
    volume = "276",
    pages = "272",
    year = "1986"
}

@article{Adams:2001sv,
    author = "Adams, A. and Polchinski, J. and Silverstein, Eva",
    title = "{Don't panic! Closed string tachyons in ALE space-times}",
    eprint = "hep-th/0108075",
    archivePrefix = "arXiv",
    reportNumber = "SLAC-PUB-8955, NSF-ITP-01-75",
    doi = "10.1088/1126-6708/2001/10/029",
    journal = "JHEP",
    volume = "10",
    pages = "029",
    year = "2001"
}

@article{Atiyah_Patodi_Singer_1975, title={Spectral asymmetry and {R}iemannian geometry. {I}}, volume={77}, DOI={10.1017/S0305004100049410}, number={1}, journal={Math. Proc. Cambridge Philos. Soc.}, author={Atiyah, M. F. and Patodi, V. K. and Singer, I. M.}, year={1975}, pages={43--69}}

@inproceedings{Witten:2019bou,
    author = "Witten, Edward and Yonekura, Kazuya",
    title = "{Anomaly Inflow and the $\eta$-Invariant}",
    booktitle = "{Memorial Volume for Shoucheng Zhang}",
    eprint = "1909.08775",
    archivePrefix = "arXiv",
    primaryClass = "hep-th",
    publisher = "WSP",
    doi = "10.1142/9789811231711_0014",
    year = "2022"
}

@article{Kaidi:2019tyf,
    author = "Kaidi, Justin and Parra-Martinez, Julio and Tachikawa, Yuji",
    title = "{Topological Superconductors on Superstring Worldsheets}",
    eprint = "1911.11780",
    archivePrefix = "arXiv",
    primaryClass = "hep-th",
    reportNumber = "IPMU-19-0164, UCLA/TEP/2019/106",
    doi = "10.21468/SciPostPhys.9.1.010",
    journal = "SciPost Phys.",
    volume = "9",
    pages = "10",
    year = "2020"
}

@article{Yonekura:2016wuc,
    author = "Yonekura, Kazuya",
    title = "{Dai-Freed theorem and topological phases of matter}",
    eprint = "1607.01873",
    archivePrefix = "arXiv",
    primaryClass = "hep-th",
    reportNumber = "IPMU-16-0094",
    doi = "10.1007/JHEP09(2016)022",
    journal = "JHEP",
    volume = "09",
    pages = "022",
    year = "2016"
}

@article{Witten:2015aba,
    author = "Witten, Edward",
    title = "{Fermion Path Integrals And Topological Phases}",
    eprint = "1508.04715",
    archivePrefix = "arXiv",
    primaryClass = "cond-mat.mes-hall",
    doi = "10.1103/RevModPhys.88.035001",
    journal = "Rev. Mod. Phys.",
    volume = "88",
    number = "3",
    pages = "035001",
    year = "2016"
}

@article{AndersonBrownPeterson1969,
  author  = {Anderson, D. W. and Brown, Jr., E. H. and Peterson, F. P.},
  title   = {Pin cobordism and related topics},
  journal = {Comment. Math. Helv.},
  volume  = {44},
  year    = {1969},
  pages   = {462--468},
  doi     = {10.1007/BF02564545}
}

@inbook{Kirby_Taylor_1991, place={Cambridge}, series={London Mathematical Society Lecture Note Series}, title={Pin structures on low-dimensional manifolds}, booktitle={Geometry of Low-Dimensional Manifolds: Symplectic Manifolds and Jones-Witten Theory}, publisher={Cambridge University Press}, author={Kirby, R.C. and Taylor, L.R.}, editor={Donaldson, S. K. and Thomas, C. B.Editors}, year={1991}, pages={177-242}, doi={10.1017/CBO9780511629341.015}, collection={London Mathematical Society Lecture Note Series}}

@book{Rudyak1998,
  author    = {Rudyak, Yuli B.},
  title     = {On Thom Spectra, Orientability, and Cobordism},
  series    = {Springer Monographs in Mathematics},
  publisher = {Springer},
  address   = {Berlin, Heidelberg},
  year      = {1998},
  doi       = {10.1007/978-3-540-77751-9},
  isbn      = {978-3-540-62043-3}
}

@book{Stong1968,
  author    = {Stong, Robert E.},
  title     = {Notes on Cobordism Theory},
  series    = {Mathematical Notes},
  volume    = {7},
  publisher = {Princeton University Press},
  address   = {Princeton, NJ},
  year      = {1968},
  doi       = {10.1515/9781400879977}
}

@book{RAY_1998,
  author    = {Kochman, Stanley O.},
  title     = {Bordism, Stable Homotopy and Adams Spectral Sequences},
  series    = {Fields Institute Monographs},
  volume    = {7},
  publisher = {American Mathematical Society},
  address   = {Providence, RI},
  year      = {1996},
  doi       = {10.1090/fim/007}
}

@article{Alvarez-Gaume:1986ghj,
    author = "Alvarez-Gaume, Luis and Ginsparg, Paul H. and Moore, Gregory W. and Vafa, Cumrun",
    title = "{An $\mathrm{O}(16) \times \mathrm{O}(16)$ Heterotic String}",
    reportNumber = "HUTP-86/A013",
    doi = "10.1016/0370-2693(86)91524-8",
    journal = "Phys. Lett. B",
    volume = "171",
    pages = "155--162",
    year = "1986"
}

@article{Fraiman:2023cpa,
    author = "Fraiman, Bernardo and Gra{\~n}a, Mariana and Parra De Freitas, H{\'e}ctor and Sethi, Savdeep",
    title = "{Non-supersymmetric heterotic strings on a circle}",
    eprint = "2307.13745",
    archivePrefix = "arXiv",
    primaryClass = "hep-th",
    doi = "10.1007/JHEP12(2024)082",
    journal = "JHEP",
    volume = "12",
    pages = "082",
    year = "2024"
}

@article{debray2024smithfibersequenceinvertible,
      title={{The Smith Fiber Sequence and Invertible Field Theories}},
      author={Debray, Arun and Devalapurkar, Sanath K. and Krulewski, Cameron and Liu, Yu Leon and Pacheco-Tallaj, Natalia and Thorngren, Ryan},
      eprint={2405.04649},
      archivePrefix={arXiv},
      primaryClass={math.AT},
      doi={10.1007/s00220-025-05505-0},
      journal={Commun. Math. Phys.},
      volume={407},
      number={2},
      pages={25},
      year={2026}
}

@article{brumfiel2018pontrjagindual3dimensionalspin,
      title={{The Pontrjagin Dual of 3-Dimensional Spin Bordism}},
      author={Greg Brumfiel and John Morgan},
      year={2018},
      eprint={1612.02860},
      archivePrefix={arXiv},
      primaryClass={math.AT},
      url={https://arxiv.org/abs/1612.02860}, 
}

@article{Kapustin:2014dxa,
    author = "Kapustin, Anton and Thorngren, Ryan and Turzillo, Alex and Wang, Zitao",
    title = "{Fermionic Symmetry Protected Topological Phases and Cobordisms}",
    eprint = "1406.7329",
    archivePrefix = "arXiv",
    primaryClass = "cond-mat.str-el",
    doi = "10.1007/JHEP12(2015)052",
    journal = "JHEP",
    volume = "12",
    pages = "052",
    year = "2015"
}

@article{Gu:2012ib,
    author = "Gu, Zheng-Cheng and Wen, Xiao-Gang",
    title = "{Symmetry-protected topological orders for interacting fermions: Fermionic topological nonlinear {\ensuremath{\sigma}} models and a special group supercohomology theory}",
    eprint = "1201.2648",
    archivePrefix = "arXiv",
    primaryClass = "cond-mat.str-el",
    doi = "10.1103/PhysRevB.90.115141",
    journal = "Phys. Rev. B",
    volume = "90",
    number = "11",
    pages = "115141",
    year = "2014"
}

@inproceedings{Distler:2009ri,
    author = "Distler, Jacques and Freed, Daniel S. and Moore, Gregory W.",
    editor = "Sati, Hisham and Schreiber, Urs",
    title = "{Orientifold Precis}",
    booktitle = "{Mathematical Foundations of Quantum Field Theory and Perturbative String Theory}",
    eprint = "0906.0795",
    archivePrefix = "arXiv",
    primaryClass = "hep-th",
    series = "Proc. Symp. Pure Math.",
    volume = "83",
    pages = "159--171",
    publisher = "Amer. Math. Soc.",
    doi = "10.1090/pspum/083/2742428",
    year = "2011"
}

@article{Debray:2023yrs,
    author = "Debray, Arun and Dierigl, Markus and Heckman, Jonathan J. and Montero, Miguel",
    title = "{The Chronicles of IIBordia: Dualities, Bordisms, and the Swampland}",
    eprint = "2302.00007",
    archivePrefix = "arXiv",
    primaryClass = "hep-th",
    reportNumber = "LMU-ASC 06/23, IFT-UAM/CSIC-23-7",
    doi = "10.4310/ATMP.241028224804",
    journal = "Adv. Theor. Math. Phys.",
    volume = "28",
    number = "3",
    pages = "805--1025",
    year = "2024"
}

@inproceedings{Witten:1985mj,
    author = "Witten, Edward",
    title = "{Global anomalies in string theory}",
    booktitle = "{Symposium on Anomalies, Geometry, Topology}",
    reportNumber = "Print-85-0620 (PRINCETON)",
    month = "6",
    year = "1985"
}

@article{Sharpe:2001bs,
    author = "Sharpe, Eric R.",
    title = "{String orbifolds and quotient stacks}",
    eprint = "hep-th/0102211",
    archivePrefix = "arXiv",
    reportNumber = "DUKE-CGTP-2001-03",
    doi = "10.1016/S0550-3213(02)00039-1",
    journal = "Nucl. Phys. B",
    volume = "627",
    pages = "445--505",
    year = "2002"
}

@article{Freed:2016rqq,
    author = "Freed, Daniel S. and Hopkins, Michael J.",
    title = "{Reflection positivity and invertible topological phases}",
    eprint = "1604.06527",
    archivePrefix = "arXiv",
    primaryClass = "hep-th",
    doi = "10.2140/gt.2021.25.1165",
    journal = "Geom. Topol.",
    volume = "25",
    pages = "1165--1330",
    year = "2021"
}

@article{Gliozzi:1976qd,
    author = "Gliozzi, F. and Scherk, Joel and Olive, David I.",
    title = "{Supersymmetry, Supergravity Theories and the Dual Spinor Model}",
    reportNumber = "CERN-TH-2253",
    doi = "10.1016/0550-3213(77)90206-1",
    journal = "Nucl. Phys. B",
    volume = "122",
    pages = "253--290",
    year = "1977"
}

@article{Tachikawa:2018njr,
    author = "Tachikawa, Yuji and Yonekura, Kazuya",
    title = "{Why are fractional charges of orientifolds compatible with Dirac quantization?}",
    eprint = "1805.02772",
    archivePrefix = "arXiv",
    primaryClass = "hep-th",
    reportNumber = "IPMU-18-0067",
    doi = "10.21468/SciPostPhys.7.5.058",
    journal = "SciPost Phys.",
    volume = "7",
    number = "5",
    pages = "058",
    year = "2019"
}

@article{Debray:2021vob,
    author = "Debray, Arun and Dierigl, Markus and Heckman, Jonathan J. and Montero, Miguel",
    title = "{The anomaly that was not meant IIB}",
    eprint = "2107.14227",
    archivePrefix = "arXiv",
    primaryClass = "hep-th",
    reportNumber = "LMU-ASC 24/21",
    doi = "10.1002/prop.202100168",
    journal = "Fortsch. Phys.",
    volume = "70",
    number = "1",
    pages = "2100168",
    year = "2022"
}

@article{Witten:1996md,
    author = "Witten, Edward",
    title = "{On flux quantization in $M$-theory and the effective action}",
    eprint = "hep-th/9609122",
    archivePrefix = "arXiv",
    reportNumber = "IASSNS-HEP-96-96",
    doi = "10.1016/S0393-0440(96)00042-3",
    journal = "J. Geom. Phys.",
    volume = "22",
    pages = "1--13",
    year = "1997"
}

@article{Witten:1985xe,
    author = "Witten, Edward",
    editor = "Salam, A. and Sezgin, E.",
    title = "{Global Gravitational Anomalies}",
    reportNumber = "PRINT-85-0246 (PRINCETON)",
    doi = "10.1007/BF01212448",
    journal = "Commun. Math. Phys.",
    volume = "100",
    pages = "197",
    year = "1985"
}

@article{Killingback:1986rd,
    author = "Killingback, T. P.",
    title = "{World Sheet Anomalies and Loop Geometry}",
    reportNumber = "PUPT-1035",
    doi = "10.1016/0550-3213(87)90229-X",
    journal = "Nucl. Phys. B",
    volume = "288",
    pages = "578",
    year = "1987"
}

@article{Freed:2019sco,
    author = "Freed, Daniel S. and Hopkins, Michael J.",
    title = "{Consistency of M-Theory on Non-Orientable Manifolds}",
    eprint = "1908.09916",
    archivePrefix = "arXiv",
    primaryClass = "hep-th",
    doi = "10.1093/qmath/haab007",
    journal = "Quart. J. Math.",
    volume = "72",
    number = "1-2",
    pages = "603--671",
    year = "2021"
}

@article{Goddard:1972iy,
    author = "Goddard, P. and Thorn, Charles B.",
    title = "{Compatibility of the Dual Pomeron with Unitarity and the Absence of Ghosts in the Dual Resonance Model}",
    doi = "10.1016/0370-2693(72)90420-0",
    journal = "Phys. Lett. B",
    volume = "40",
    pages = "235--238",
    year = "1972"
}

@article{Banks:1988yz,
    author = "Banks, Tom and Dixon, Lance J.",
    title = "{Constraints on String Vacua with Space-Time Supersymmetry}",
    reportNumber = "PUPT-1086, SCIPP-8805",
    doi = "10.1016/0550-3213(88)90523-8",
    journal = "Nucl. Phys. B",
    volume = "307",
    pages = "93--108",
    year = "1988"
}

@article{Brown1972Kervaire,
  author  = {Brown, Jr., Edgar H.},
  title   = {Generalizations of the {K}ervaire invariant},
  journal = {Ann. Math.},
  volume  = {95},
  year    = {1972},
  pages   = {368--383},
  doi     = {10.2307/1970804}
}

@Inbook{Atiyah1970,
author="Atiyah, Michael
and Hirzebruch, Friedrich",
title="{Spin-Manifolds and Group Actions}",
bookTitle="Essays on Topology and Related Topics: Memoires d{\'e}di{\'e}s {\`a} Georges de Rham",
year="1970",
publisher="Springer Berlin Heidelberg",
address="Berlin, Heidelberg",
pages="18--28",
isbn="978-3-642-49197-9",
doi="10.1007/978-3-642-49197-9_3",
url="https://doi.org/10.1007/978-3-642-49197-9_3"
}

@article{Diaconescu:2000wy,
    author = "Diaconescu, Duiliu-Emanuel and Moore, Gregory W. and Witten, Edward",
    title = "{$E_8$ gauge theory, and a derivation of $K$-theory from $M$-theory}",
    eprint = "hep-th/0005090",
    archivePrefix = "arXiv",
    reportNumber = "IASSNS-HEP-00-39",
    doi = "10.4310/ATMP.2002.v6.n6.a2",
    journal = "Adv. Theor. Math. Phys.",
    volume = "6",
    pages = "1031--1134",
    year = "2003"
}

@article{Sati:2011rw,
    author = "Sati, Hisham",
    title = "{Twisted topological structures related to M-branes II: Twisted Wu and Wu$^c$ structures}",
    eprint = "1109.4461",
    archivePrefix = "arXiv",
    primaryClass = "hep-th",
    doi = "10.1142/S0219887812500569",
    journal = "Int. J. Geom. Meth. Mod. Phys.",
    volume = "09",
    pages = "1250056",
    year = "2012"
}

@article{Distler:2010an,
    author = "Distler, Jacques and Freed, Daniel S. and Moore, Gregory W.",
    title = "{Spin structures and superstrings}",
    eprint = "1007.4581",
    archivePrefix = "arXiv",
    primaryClass = "hep-th",
    reportNumber = "UTTG-07-10",
    doi = "10.4310/SDG.2010.v15.n1.a4",
    journal = "Surv. Differ. Geom.",
    volume = "15",
    pages = "99--130",
    year = "2011"
}

@article{Cheng:2023owv,
    author = "Cheng, Peng and Melnikov, Ilarion V. and Minasian, Ruben",
    title = "{Flat F-theory and friends}",
    eprint = "2306.00865",
    archivePrefix = "arXiv",
    primaryClass = "hep-th",
    doi = "10.1007/JHEP01(2024)027",
    journal = "JHEP",
    volume = "01",
    pages = "027",
    year = "2024"
}

@article{Adams:2005rb,
    author = "Adams, A. and Liu, X. and McGreevy, J. and Saltman, A. and Silverstein, Eva",
    title = "{Things fall apart: Topology change from winding tachyons}",
    eprint = "hep-th/0502021",
    archivePrefix = "arXiv",
    reportNumber = "SLAC-PUB-11011, SU-ITP-05-06, HUTP-05-A0006",
    doi = "10.1088/1126-6708/2005/10/033",
    journal = "JHEP",
    volume = "10",
    pages = "033",
    year = "2005"
}

@article{Kaidi:2019pzj,
    author = "Kaidi, Justin and Parra-Martinez, Julio and Tachikawa, Yuji",
    title = "{Classification of String Theories via Topological Phases}",
    eprint = "1908.04805",
    archivePrefix = "arXiv",
    primaryClass = "hep-th",
    doi = "10.1103/PhysRevLett.124.121601",
    journal = "Phys. Rev. Lett.",
    volume = "124",
    number = "12",
    pages = "121601",
    year = "2020"
}

@article{Bhardwaj:2016clt,
    author = "Bhardwaj, Lakshya and Gaiotto, Davide and Kapustin, Anton",
    title = "{State sum constructions of spin-TFTs and string net constructions of fermionic phases of matter}",
    eprint = "1605.01640",
    archivePrefix = "arXiv",
    primaryClass = "cond-mat.str-el",
    doi = "10.1007/JHEP04(2017)096",
    journal = "JHEP",
    volume = "04",
    pages = "096",
    year = "2017"
}

@article{Atick:1988si,
    author = "Atick, Joseph J. and Witten, Edward",
    title = "{The Hagedorn Transition and the Number of Degrees of Freedom of String Theory}",
    reportNumber = "IASSNS-HEP-88-14",
    doi = "10.1016/0550-3213(88)90151-4",
    journal = "Nucl. Phys. B",
    volume = "310",
    pages = "291--334",
    year = "1988"
}

@article{Acharya:2020hsc,
    author = "Acharya, Bobby Samir and Aldazabal, Gerardo and Andr{\'e}s, Eduardo and Font, Anamar{\'\i}a and Narain, Kumar and Zadeh, Ida G.",
    title = "{Stringy Tachyonic Instabilities of Non-Supersymmetric Ricci Flat Backgrounds}",
    eprint = "2010.02933",
    archivePrefix = "arXiv",
    primaryClass = "hep-th",
    doi = "10.1007/JHEP04(2021)026",
    journal = "JHEP",
    volume = "04",
    pages = "026",
    year = "2021"
}

@article{Witten:2023snr,
    author = "Witten, Edward",
    title = "{Anomalies and Nonsupersymmetric D-Branes}",
    eprint = "2305.01012",
    archivePrefix = "arXiv",
    primaryClass = "hep-th",
    month = "5",
    year = "2023"
}

@article{Jensen:2017eof,
    author = "Jensen, Kristan and Shaverin, Evgeny and Yarom, Amos",
    title = "{{\textquoteright}t Hooft anomalies and boundaries}",
    eprint = "1710.07299",
    archivePrefix = "arXiv",
    primaryClass = "hep-th",
    doi = "10.1007/JHEP01(2018)085",
    journal = "JHEP",
    volume = "01",
    pages = "085",
    year = "2018"
}

@book{Polchinski:1998rr,
    author = "Polchinski, J.",
    title = "{String theory. Vol. 2: Superstring theory and beyond}",
    doi = "10.1017/CBO9780511618123",
    isbn = "978-0-511-25228-0, 978-0-521-63304-8, 978-0-521-67228-3",
    publisher = "Cambridge University Press",
    series = "Cambridge Monographs on Mathematical Physics",
    month = "12",
    year = "2007"
}

@article{Gepner:1987qi,
    author = "Gepner, Doron",
    editor = "Schellekens, B.",
    title = "{Space-Time Supersymmetry in Compactified String Theory and Superconformal Models}",
    reportNumber = "Print-87-0370 (PRINCETON), PUPT-1056",
    doi = "10.1016/0550-3213(88)90397-5",
    journal = "Nucl. Phys. B",
    volume = "296",
    pages = "757",
    year = "1988"
}

@article{Gepner:1987vz,
    author = "Gepner, Doron",
    title = "{Exactly Solvable String Compactifications on Manifolds of $\mathrm{SU}(N)$ Holonomy}",
    reportNumber = "PRINT-87-0727 (PRINCETON), PUPT-1066",
    doi = "10.1016/0370-2693(87)90938-5",
    journal = "Phys. Lett. B",
    volume = "199",
    pages = "380--388",
    year = "1987"
}

@article{Banks:1987cy,
    author = "Banks, Tom and Dixon, Lance J. and Friedan, Daniel and Martinec, Emil J.",
    title = "{Phenomenology and Conformal Field Theory Or Can String Theory Predict the Weak Mixing Angle?}",
    reportNumber = "SLAC-PUB-4377",
    doi = "10.1016/0550-3213(88)90551-2",
    journal = "Nucl. Phys. B",
    volume = "299",
    pages = "613--626",
    year = "1988"
}

@article{Tachikawa:2021mby,
    author = "Tachikawa, Yuji and Yamashita, Mayuko",
    title = "{Topological Modular Forms and the Absence of All Heterotic Global Anomalies}",
    eprint = "2108.13542",
    archivePrefix = "arXiv",
    primaryClass = "hep-th",
    doi = "10.1007/s00220-023-04761-2",
    journal = "Commun. Math. Phys.",
    volume = "402",
    number = "2",
    pages = "1585--1620",
    year = "2023",
    note = "[Erratum: Commun.Math.Phys. 402, 2131 (2023)]"
}

@incollection{gilkey,
    author = "Gilkey, Peter B. and Botvinnik, Boris",
    title = "{The eta invariant and the equivariant spin bordism of spherical space form 2 groups}",
    booktitle = "{New Developments in Differential Geometry}",
    series = "Math. Appl.",
    volume = "350",
    pages = "213--223",
    publisher = "Kluwer",
    year = "1996",
    doi = "10.1007/978-94-009-0149-0_16"
}
\inputencoding{utf8}

\end{document}